\documentclass[%
reprint,
superscriptaddress,
 amsmath,amssymb,
 aps
]{revtex4-2}

\usepackage[utf8]{inputenc}
\usepackage{graphicx}%
\usepackage{multirow}%
\usepackage{amsmath,amssymb,amsfonts}%
\usepackage{amsthm}%
\usepackage{mathrsfs}%
\usepackage[title]{appendix}%
\usepackage{xcolor}%
\usepackage{textcomp}%
\usepackage{booktabs}%
\usepackage{listings}%
\usepackage{indentfirst}
\usepackage{float}
\usepackage{hyperref}

\usepackage[all]{hypcap}
\hypersetup{
    colorlinks=true,
    linkcolor=blue,
    filecolor=blue,      
    urlcolor=blue,
    citecolor=blue
    }

\usepackage{lineno} 
\usepackage{soul} 

\let\up\uparrow
\let\down\downarrow

\newcommand{\beq}{\begin{equation}}
\newcommand{\eeq}{\end{equation}}
\newcommand{\beqa}{\begin{eqnarray}}
\newcommand{\eeqa}{\end{eqnarray}}

\usepackage{titlesec}
\titleformat*{\section}{\centering\footnotesize\textbfseries\uppercase}

\usepackage[paperwidth=210mm,paperheight=297mm,centering,hmargin=1.7cm,vmargin=1.5cm]{geometry}

\begin{document}

\nolinenumbers

\title{\large \textbf{Probing the Fermi Sea Topology in a Quantum Gas}}


\author{Cyprien Daix}
\affiliation{Laboratoire Kastler Brossel, ENS-Universit\'{e} PSL, CNRS, Sorbonne Universit\'{e}, Coll\`{e}ge de France, 24 rue Lhomond, 75005, Paris, France}
\author{Pok Man Tam}
\affiliation{Princeton Center for Theoretical Science, Princeton University, Princeton, New Jersey 08544, USA}
\author{Maxime Dixmerias}
\affiliation{Laboratoire Kastler Brossel, ENS-Universit\'{e} PSL, CNRS, Sorbonne Universit\'{e}, Coll\`{e}ge de France, 24 rue Lhomond, 75005, Paris, France}
\author{Joris Verstraten}
\affiliation{Laboratoire Kastler Brossel, ENS-Universit\'{e} PSL, CNRS, Sorbonne Universit\'{e}, Coll\`{e}ge de France, 24 rue Lhomond, 75005, Paris, France}
\author{Tim de Jongh}
\thanks{Present address: JILA, National Institute of Standards and Technology, and Department of Physics, University of Colorado, Boulder, CO 80309, USA}
\affiliation{Laboratoire Kastler Brossel, ENS-Universit\'{e} PSL, CNRS, Sorbonne Universit\'{e}, Coll\`{e}ge de France, 24 rue Lhomond, 75005, Paris, France}
\author{\\Bruno Peaudecerf}
\affiliation{Laboratoire Collisions Agr\'egats R\'eactivit\'e, UMR 5589, FERMI, UT3, Universit\'e de Toulouse, CNRS, 118 Route de Narbonne, 31062, Toulouse CEDEX 09, France}
\author{Charles L. Kane}
\affiliation{Department of Physics and Astronomy, University of Pennsylvania, Philadelphia, Pennsylvania 19104, USA}
\author{Tarik Yefsah}
\thanks{Correspondence to be addressed to: \href{mailto:tarik.yefsah@lkb.ens.fr}{tarik.yefsah@lkb.ens.fr}}
\affiliation{Laboratoire Kastler Brossel, ENS-Universit\'{e} PSL, CNRS, Sorbonne Universit\'{e}, Coll\`{e}ge de France, 24 rue Lhomond, 75005, Paris, France}

\date{\today}

\begin{abstract}
\quad Pauli's exclusion principle forces fermions to occupy distinct quantum states, creating  a filled region of momentum space at low temperature, the Fermi sea, whose topology governs the system's response to perturbations and the nature of its correlation functions. Recent theory predicts that for non-interacting fermions, the Euler characteristic of a $D$-dimensional Fermi sea -- the topological invariant that describes its shape -- is encoded in its ($D$+1)-point density correlations. Here we experimentally demonstrate this connection in a two-dimensional degenerate gas of neutral $^{6}$Li atoms using single-atom-resolved imaging. By measuring three- and four-point connected density correlations in real space, we directly extract topological invariants of the underlying Fermi sea, including the Euler characteristic. Our results are in remarkable agreement with ideal-gas predictions, despite the presence of sizeable interactions, and establish a new pathway for probing many-body topology through correlation measurements.
\end{abstract}

\maketitle

\subsection*{Introduction}

The symmetrization postulate of quantum mechanics is a cornerstone of many-body physics, assigning all fundamental particles to one of two possible families: fermions and bosons. For fermions such as electrons, neutrons, and protons -- the constituents of all ordinary matter around us -- it requires that their wavefunction be antisymmetric, generalizing the Pauli exclusion principle originally proposed for electrons, with profound implications. This principle underlies the structure and stability of matter, and is at the origin of a wide range of phenomena in many-body systems, from magnetism to the formation of Cooper pairs in superconductors and the superfluidity of neutron stars.

Pauli's exclusion principle gives rise to another key object for the description of such collective phenomena: the Fermi sea -- the filled region of momentum space with unity occupation per state at zero temperature. Although a direct consequence of the antisymmetrization of the fermionic wavefunction, the Fermi sea is an important concept in its own right:  key material properties find their origin in its attributes. In addition to geometrical properties, such as size and curvature, the Fermi sea is a {\it topological} object that is characterized by an integer topological invariant called the Euler characteristic $\chi_{\rm F}$.


\begin{figure}[t!]
	\centering
	\includegraphics[width = \columnwidth]{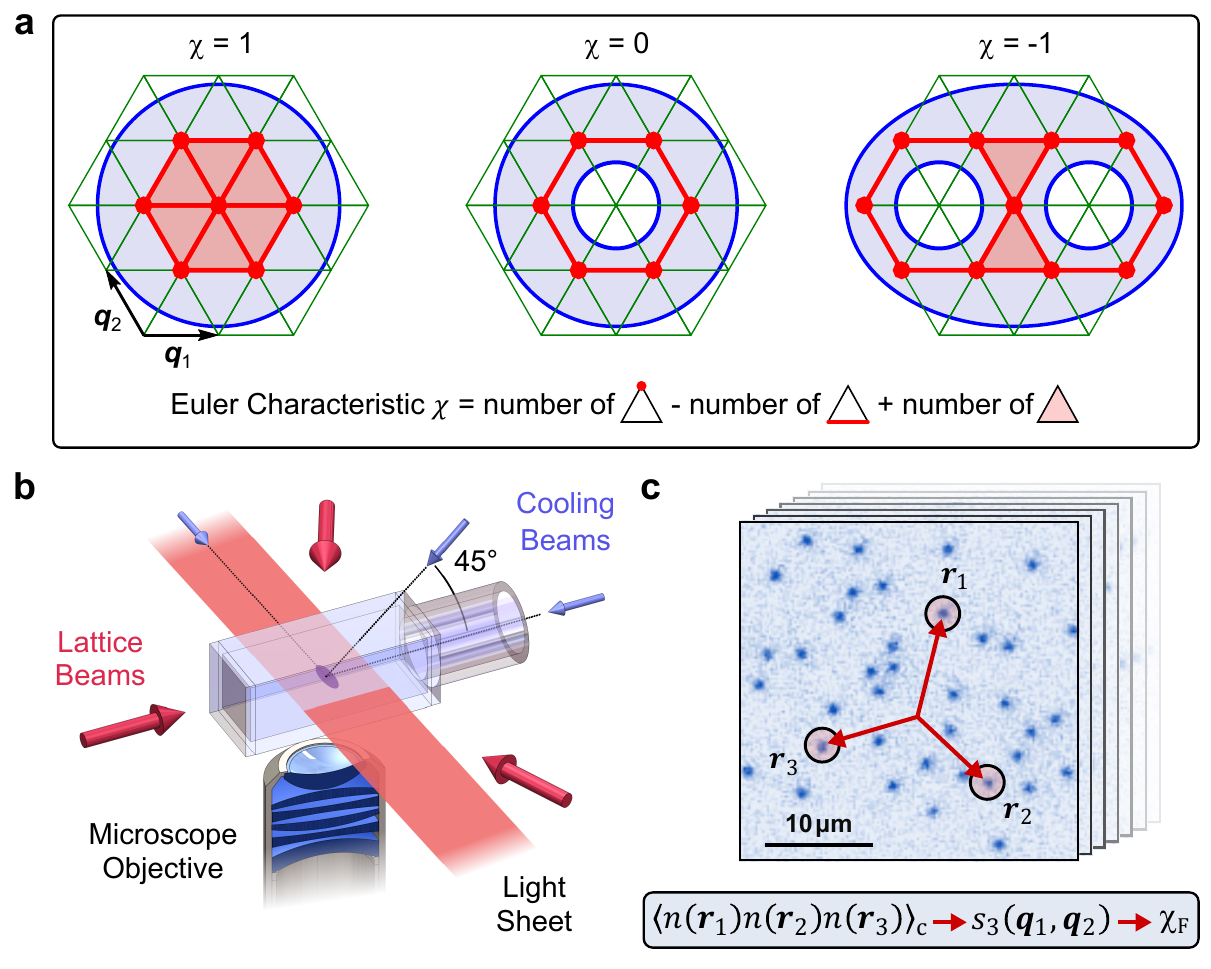}
	\caption{\textbf{Probing the Fermi Sea Topology via Single-Atom Imaging.} 
    (a)  The Euler characteristic $\chi$ is an integer that distinguishes topologically distinct shapes.  Originally formulated by Euler to characterize polyhedra (like a cube or a tetrahedron, with $\chi=2$), $\chi$ can be determined by counting the number of points, links and faces. Applied to two dimensional regions (blue), $\chi$ can be evaluated by introducing a triangulation (green mesh), defined by vectors ${\bf q}_{1}$ and ${\bf q}_{2}$, and computing $\chi$ from the triangles, lines and points included in the region (red).   For a sufficiently fine mesh, $\chi$ is independent of the triangulation.  
    (b) A dilute Fermi gas is held in a highly oblate gaussian trap (``light sheet") providing a quasi-two-dimensional geometry. A combination of a 2D pinning lattice (red arrows) and laser cooling (blue arrows) produces images of the spatial distribution of atoms with single atom resolution, as shown in 
    (c)  The three-point density correlations $\langle n({\bf r}_1)n({\bf r}_2)n({\bf r}_3)\rangle_{\rm c}$ in a Fermi gas are extracted from  averages over quantum gas microscopy images. The Euler characteristic of the Fermi sea, $\chi_{\rm F}$, is encoded in the Fourier transform $s_3({\bf q}_1,{\bf q}_2)$, where ${\bf q}_{1,2}$ define the triangulation in (a). 
}
	\label{fig:fig1}
\end{figure}


\begin{figure*}[t!]
	\centering
	\includegraphics[width = 0.8\textwidth]{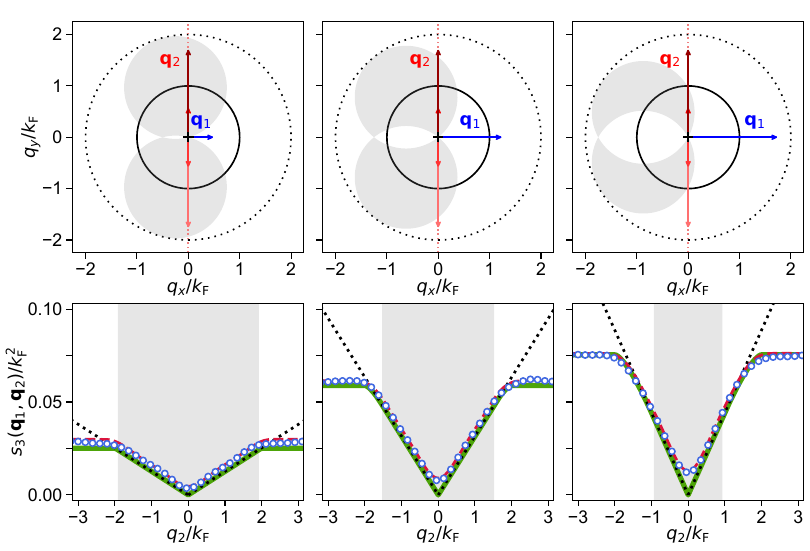}
	\caption{\textbf{Measurement of Euler characteristic from density correlations.} $s_3(\textbf{q}_1, \textbf{q}_2)$ for different configurations of $\textbf{q}_1$ and $\textbf{q}_2$ shown in the top row, with $\textbf{q}_2 \perp \textbf{q}_1$. For each column, $\textbf{q}_1$ is kept fixed, while $\textbf{q}_2$ is varied along the dotted red line. The Fermi surface is pictured as a solid black circle, with a circle of radius $2k_{\rm F}$ also shown as a dotted black curve. The shaded area corresponds to the region of validity of the topological formula, $S_{\rm topo}$ for the chosen value of $\textbf{q}_1$, as defined by Eq.~\eqref{eq:validity_criterion}.  From left to right: $|\textbf{q}_1|/k_{\rm F} = 0.5$, $1.25$, $1.75$. Bottom row:
    Measured values of $s_3$ as a function of the amplitude $q_2$, averaged over different absolute orientations of $(\textbf{q}_1, \textbf{q}_2)$ (blue markers). The data is compared to the topological formula Eq.~\eqref{eq:topological_formula} with $\chi_{\rm F}\, =\, 1$ (black dotted line), the exact analytical formula Eq.~(\ref{eq:exact_formula_T0}) (green solid line), and finite temperature numerical predictions (red dash-dotted line). Within $S_{\rm topo}$, indicated by the shaded areas, $s_3$ has the characteristic V-shape predicted by the topological formula and is proportional to $\chi_{\rm F} |\textbf{q}_1||\textbf{q}_2|$. Errorbars show the standard error of the mean and if not visible are smaller than the markers.}
	\label{fig:fig2}
\end{figure*}


In two dimensional (2D) ballistic metals, $\chi_{\rm F}$ is associated with a quantized nonlinear conductance \cite{kane2022}.    This generalizes the quantized Landauer conductance in one dimension (1D), and highlights the topological origin of the quantized steps in the linear conductance observed in quantum point contacts \cite{vanwees1988}, 1D semiconductors \cite{honda1995, vanweperen2013}, carbon nanotubes \cite{frank1998} and quantum gases \cite{krinner2015}. This behavior is reminiscent of the integer quantized Hall effect \cite{klitzing1980} in which topological properties of Bloch bands lead to quantized plateaus in the Hall conductivity \cite{thouless1982}. In ultracold Fermi gases, properties have been observed that reflect geometric features of the underlying Fermi sea, including a suppression of spontaneous \cite{sanner2021, deb2021, margalit2021} and stimulated \cite{jannin2022} emission and the formation of Cooper-like pairing in the presence of attractive interactions \cite{zwierlein2013,holten_2022}.  

However, until now, a direct measurement of the Euler characteristic of the Fermi sea beyond the 1D case has never been performed.  In recent years, several theoretical pathways to accomplish this have been proposed.  These include measuring quantized nonlinear response functions \cite{kane2022, yang2022, zhang2023}, Andreev state transport \cite{tam2023, tam2023a}, multipartite number fluctuations \cite{tam2022} and correlation functions \cite{tam2024}. In particular the latter prediction has shown that this information is encoded in equal-time, higher-order density correlations, implying that topological properties of the Fermi gas can be extracted directly from equilibrium particle configurations, without applying fields or driving currents.

Here we realize this idea experimentally using a degenerate gas of neutral Lithium-6 ($^6$Li) atoms confined to two dimensions. Using in-situ single-atom imaging, we measure for each spin-component three-point and four-point connected density correlations of the system and directly reveal the Fermi sea topology via the measurement of topological invariants, extracting the Euler characteristic which we find to agree with the expected value $\chi_{\rm F}=1$ for the circular Fermi sea of a single-component ideal Fermi gas. Remarkably, our results are obtained in conditions that deviate significantly from those under which the theory was originally derived, including the presence of relatively strong attractive inter-spin interactions and the absence of translational symmetry of the system. This strikingly highlights both the distinctive robustness of the Euler characteristic as a topological invariant and the unexpected range of applicability of this universal property. Our results experimentally demonstrate that equal-time correlation functions can reveal topological invariants in equilibrium many-body systems, establishing a new link between topology and correlations in quantum matter.

\subsection*{Topology of the Fermi Sea}


In a translation-invariant ideal 2D Fermi gas at zero temperature, the Euler characteristic is encoded in the three-point connected correlation function $\langle \hat{n}(\textbf{0}) \hat{n}(\textbf{r}_1) \hat{n}(\textbf{r}_2) \rangle_{\rm c}$, where $\hat{n}$ denotes the density operator. Taking the Fourier transform:

\beq
s_3 (\textbf{q}_1,\,\textbf{q}_2)\equiv\int d\textbf{r}_1 d\textbf{r}_2 e^{-i\textbf{q}_1\cdot\textbf{r}_1-i\textbf{q}_2\cdot\textbf{r}_2}\langle \hat{n}(\textbf{0}) \hat{n}(\textbf{r}_1) \hat{n}(\textbf{r}_2) \rangle _{\rm c}
\eeq and examining its behavior for sufficiently small values of $\mathbf{q_1}$ and $\mathbf{q_2}$ (a condition that we clarify below) reveals $\chi_{\rm F}$ in a remarkably simple way:
\beq
s_3(\textbf{q}_1,\,\textbf{q}_2) = \frac{V_2(\textbf{q}_1,\,\textbf{q}_2)}{(2\pi)^2}\chi_{\rm F},
\label{eq:topological_formula}
\eeq
where $V_2(\textbf{q}_1,\,\textbf{q}_2) = \vert \textbf{q}_1 \times \textbf{q}_2 \vert$ is the area of the parallelogram formed by $\{\textbf{q}_1,\,\textbf{q}_2\}$. This result has a clear topological origin that was elucidated in Ref.~\cite{tam2024}: for a given triangulation of momentum space by \mbox{$\{\textbf{q}_a\}_{a = 1, 2, 3}\equiv\{{\textbf{q}_1, \textbf{q}_2, \textbf{q}_3}=-\textbf{q}_1 - \textbf{q}_2\}$}, resulting in $P$ number of points, $L$ number of links and $F$ number of faces inside the Fermi sea, the algebraic structure of $s_3$ is proportional to $P-L+F$, see Fig.~\ref{fig:fig1}a, which is reminiscent of Euler's characterization of polyhedra \cite{euler1758}. Similarly, one can show that the Fourier transform of the $M$-th order connected correlation function, $s_M$, vanishes for all $M \ge 4$ for small enough $\{\textbf{q}_a\}$, which provides another distinctive topological signature. These results are universal as they apply to any shape of the Fermi surface and are generalizable to any dimension. In a $D$-dimensional Fermi gas, the Euler characteristic is reflected in the ($D$+1)--point equal-time density-density correlations, with the $D$-dimensional parallelepiped volume $V_D(\textbf{q}_1,\,\textbf{q}_2, ...,\,\textbf{q}_{M-1})/(2\pi)^D$ appearing as the geometric prefactor in Eq.~\eqref{eq:topological_formula} \cite{tam2024}. Likewise, $s_M = 0$ for all $M \ge D + 2$. In the following, we probe these properties experimentally and validate this universal prediction in a 2D atomic Fermi gas. 

\subsection*{In-situ Single-Atom Imaging of Fermi Gases}

Our experiment starts with a two-spin mixture ($\uparrow$ and $\downarrow$) of $^6$Li atoms, prepared with balanced populations $\langle N_{\up} \rangle \approx \langle N_{\down} \rangle \approx 70$, and confined to a single plane by a laser-induced trap (see Fig.~\ref{fig:fig1}b) that provides strong harmonic confinement along the vertical $z$-direction with frequency $\omega_z = 2\pi \times 1.125(50)\,$kHz. In the $xy$-plane, the atoms experience a nearly harmonic potential, which is sufficiently shallow to ignore quantization of its energy levels away from the edges \cite{castin2007}, leading to a continuously decreasing atomic density as one moves away from the center.
We will focus our analysis on a central quasi-homogeneous region of the cloud. Denoting $\bar{n}$ the average density of a single component over this region, we define a characteristic Fermi wavenumber $k_{\rm F}\equiv \sqrt{4\pi \bar{n}}$, Fermi energy $E_{\rm F} \equiv \hbar^2k_{\rm F}^2/(2m)$, and Fermi temperature $T_{\rm F} = E_{\rm F}/ k_{\rm B}$, where $\hbar$ is the reduced Planck constant, $m$ the atomic mass and $k_{\rm B}$ the Boltzmann constant. 

Our samples are prepared at finite temperature, with reduced temperatures $T/T_{\rm F} \approx 0.1$ to $0.2$, and feature sizeable contact interactions. Interactions are isotropic, attractive, and occur exclusively between atoms in different spin states.
The interaction strength is characterized by the dimensionless parameter ${\cal I}= -1/\log(k_{\rm F}a)$, where the two-dimensional scattering length $a$ is tunable via a magnetic field near a Feshbach resonance (see Methods). At sufficiently low temperature the system is superfluid for any nonzero value of ${\cal I}$ \cite{levinsen2015}. The weakly attractive regime ${\cal I} \rightarrow 0^-$ corresponds to a Bardeen-Cooper-Schrieffer (BCS) state, and the system is considered strongly interacting when $\vert{\cal I}\vert\gtrsim1$. In this work, we study samples with ${\cal I}= -0.49(2)$,  $-0.27(1)$, and ${\cal I}=-0.129(3)$ (see Ref. \cite{daix2025} for details on the preparation). For clarity, the data presented in the main text focuses on the case ${\cal I}= -0.129(3)$, while the other cases are presented in the Supplementary materials~\cite{supmat}.

We probe each spin component of the cloud in situ, at the single atom level and with a resolution well below the inter-particle spacing, using continuum quantum gas microscopy \cite{verstraten2025,dejongh2025,yao2025,xiang2025}. This technique involves freezing the motion of atoms initially evolving in continuous space by rapidly switching on a deep optical lattice and then illuminating the atoms with cooling light to induce fluorescence while holding them in individual lattice sites \cite{verstraten2025,dejongh2025}. A typical experimental image is shown in Fig.~\ref{fig:fig1}c, where each atom is resolved with $>99.5\%$ fidelity, providing pristine access to the system's density-density correlations up to large distances and with a high degree of precision \cite{dejongh2025,daix2025}. For each interaction strength, we acquire on the order of 1400 images (700 per spin state), from which we evaluate three-point and four-point connected density-density correlations in real space.


\begin{figure}[t!]
\centering
	\includegraphics[width = \columnwidth]{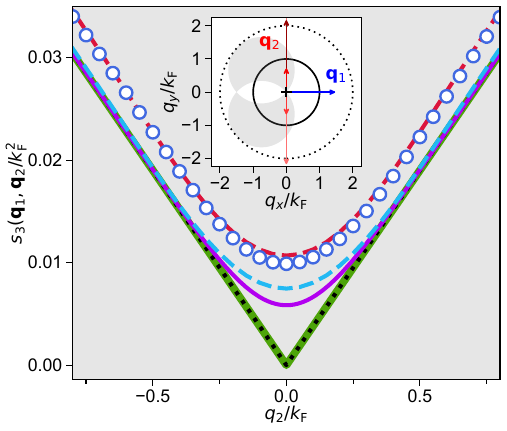}
	\caption{\textbf{Robustness of topological scaling.} Zoom on the central part of $s_3$ for small values of $q_{2}$. The markers are experimental data. Solid green line: analytical prediction at $T=0$, Eq.~(\ref{eq:exact_formula_T0}); dotted black line: topological formula Eq.~(\ref{eq:topological_formula}); solid purple line: finite temperature prediction Eq.~(\ref{eq:finiteT_expansion}); cyan dashed line: numerical results for a pure 2D system, which incorporate finite-size effects; red dashed line: numerical results including the effect of population of excited transverse motional levels. $S_{\rm topo}$ is shown as a shaded area. The chosen configuration for the triangulating vectors is shown as an inset, corresponding to $|\textbf{q}_1|/k_{\rm F} = 1.5$. Finite temperature and finite-size effects both contribute to smoothing the singularity predicted at $q_{2}=0$. The offset between the experimental data points and the linear branches of the topological formula is fully explained by the presence of atoms in excited $z$-levels.}
	\label{fig:fig3}
\end{figure}



\begin{figure*}[!t]
	\centering
	\includegraphics[width = \textwidth]{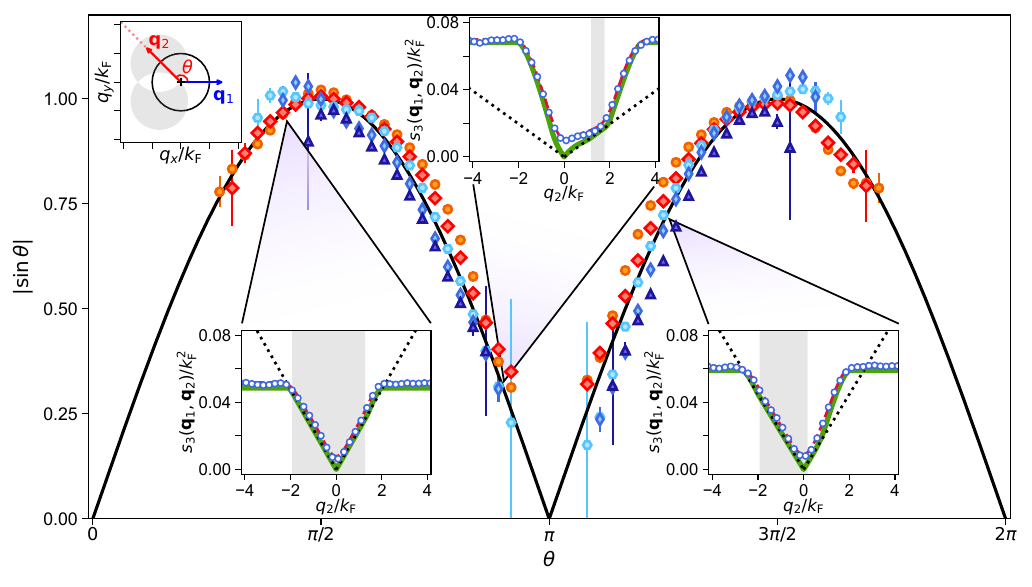}
	\caption{\textbf{Volume scaling of $s_3$.} Measured slope of $s_3$ (times $(2\pi)^2/|\textbf{q}_1|$) within the topological region for different values of the angle $\theta$ between $\textbf{q}_1$ and $\textbf{q}_2$ (top left inset) and different values of $|\textbf{q}_1|$. The orange circles, red squares, light blue hexagons, blue diamonds and dark blue triangles correspond to $|\textbf{q}_1|/k_{\rm F} = 0.75,\, 1,\, 1.25,\, 1.5,\, 1.75$ respectively. The prediction of the topological formula is shown in black. In insets, we show corresponding $s_3$ for different values of $\theta \mod \pi$. The markers are experimental data. Solid green line: analytical prediction at $T=0$, Eq.~(\ref{eq:exact_formula_T0}); dotted black line: topological formula, Eq.~(\ref{eq:topological_formula}); dashed red line: numerical results. $S_{\rm topo}$ is shown as a shaded area.}
	\label{fig:fig4}
\end{figure*}


\subsection*{Measuring the Euler Characteristic}

We start our analysis with the three-point correlations $\langle \hat{n}(\textbf{0}) \hat{n}(\textbf{r}_1) \hat{n}(\textbf{r}_2) \rangle_{\rm c}$, to which we apply a discrete Fourier transformation to obtain $s_3$ for various triangulations by the vectors $\{\textbf{q}_a\}_{a=1,2,3}$. Since the vectors are in a plane, the lengths of the vectors $\textbf{q}_1$ and $\textbf{q}_2$, and the angle $\theta$ between them are the only independent parameters. We first consider the case where $\textbf{q}_1$ and $\textbf{q}_2$ are orthogonal, which is a favorable configuration for the validity of Eq.\,\eqref{eq:topological_formula}, set by the condition (see~\cite{supmat}):
\beq
R_{\{{\bf q}\}}\equiv \frac{|{\bf q}_1||{\bf q}_2||{\bf q}_3|}{2|{\bf q}_1\times {\bf q}_2|}<k_{\rm F}.
\label{eq:validity_criterion}
\eeq 
As defined, $R_{\{{\bf q}\}}$ is the circumscribing radius for the triangle formed by $\{\textbf{q}_a\}_{a = 1, 2, 3}$, which is the elementary region for the triangulation of the Fermi sea as depicted in Fig.~\ref{fig:fig1}a. 
Within the momentum space region $S_{\rm topo}$ defined by this criterion, Eq.~\eqref{eq:topological_formula} is exact at $T=0$. In Fig.~\ref{fig:fig2}, we show examples of experimentally measured $s_3$ for different configurations, where $\textbf{q}_1$ is kept fixed and we vary the amplitude $q_2$. By averaging over different absolute angles of $\textbf{q}_1$, we obtain measurements of $s_3$ with an excellent signal-to-noise ratio and observe the distinctive V-shape expected from Eq.~\eqref{eq:topological_formula}. Within most of its validity region (shown as a grey area) we find excellent quantitative agreement with the slope predicted by Eq.~\eqref{eq:topological_formula} and the expected value $\chi_{\rm F} = 1$, with no fitting parameters. For instance, in the representative cases shown in Fig.~\ref{fig:fig2}, a linear fit yields $\chi_{\rm F}=1.015(7)$, $0.985(6)$, and $0.96(2)$ from left to right, respectively. The stated error corresponds to the statistical error of the fit. In addition, we expect small systematic shifts of the order of $\pm0.01$ (see~\cite{supmat}).
The expected kink of $s_3$ near $q_2=0$ from Eq.~\eqref{eq:topological_formula} appears rounded, which is a combined effect of the finite size and finite temperature of the system, which we quantitatively address in the next section. These results represent a direct observation of the Fermi sea topology. 

To interpret our experimental results further, we have derived an exact analytical expression valid at $T=0$ for a translationally invariant system that extends beyond the criterion in Eq.\,\eqref{eq:validity_criterion}, which reads 
\begin{equation}
    s_3 =
    \begin{cases}
    \frac{1}{(2\pi)^2}|{\bf q}_1 \times {\bf q}_2|,&\text{if $R_{\{{\bf q}\}} \leq k_{\rm F}$} \\
        \frac{k_{\rm F}^2}{4\pi}\left(1 + \sum_{a=1}^3 \sigma_a G(\frac{|{\bf q}_a|}{2k_{\rm F}}) \right), &\text{if $R_{\{{\bf q}\}} \geq k_{\rm F}$}
    \end{cases}.
    \label{eq:exact_formula_T0}
\end{equation}
In the above we have defined $\sigma_{a} = {\rm sgn}[{\bf q}_b\cdot{\bf q}_c]$ for $a\ne b\ne c = 1,2,3$, $G(X) =  \frac{2}{\pi}\left(\cos^{-1} X - X\sqrt{1-X^2}\right)\Theta(1-X)$ with $\Theta(X)$ the Heaviside step function (see~\cite{supmat} for a derivation).
This prediction, shown as a solid green line in Fig.~\ref{fig:fig2}, reproduces the observed behavior of $s_3$ in and outside $S_{\rm topo}$ with a remarkable accuracy, except for a small vertical shift of the data and a smoothing of the singularity, which we explain below.

\subsection*{Topological Robustness}


\begin{figure}[!t]
\centering
	\includegraphics[width = \columnwidth]{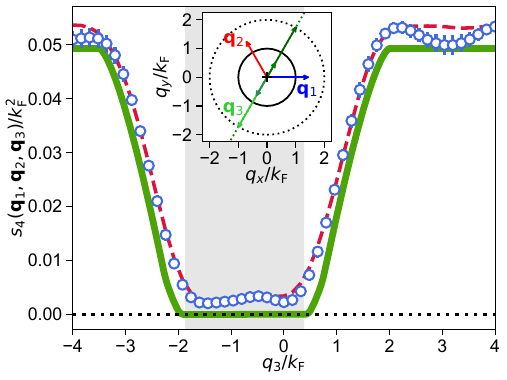}
	\caption{\textbf{Four-point connected density correlations.} $s_4$ for $|{\bf q}_1|/k_{\rm F}=1.5$, $|{\bf q}_2|/k_{\rm F}=1.5$ with $\theta_{12}=2\pi/3$ and $\theta_{13}=\theta_{12}/2\ \rm{mod}\ \pi$. Blue markers: experimental data; green curve: analytical result; red curve: numerical results. $s_4$ is seen to remain nearly constant and close to zero over a large region in momentum space, which coincides with the region of validity for $s_4=0$ (grey area). (Inset) Plot of the chosen configuration $\{\textbf{q}_p \}_{p=1,2,3}$.}
	\label{fig:fig5}
\end{figure}


The clarity and precision with which the Fermi sea topology can be extracted from our measurements is striking, given all the deviations from ideal conditions present in our system: (I) the temperature is finite, (II) we probe a finite system (see discussion below)
(III) the gas is not strictly 2D, with a small fraction of atoms occupying the first excited $z$-level, (IV) in-plane translational symmetry is broken due to the trapping potential, and (V) the system is in the interacting regime, with sizeable inter-spin interactions.  

To assess the grounds for such robustness, we first investigate quantitatively  the impact of alterations (I)–(III) by comparing our results with an analytical prediction at finite temperature (see Eq.~\eqref{eq:finiteT_expansion} below) and with numerical calculations for homogeneous ideal Fermi gases at finite temperature~\cite{supmat}. The numerical calculations are performed either in 2D or in quasi-2D geometry, taking into account the vertical degrees of freedom, and employ the same discrete Fourier transform algorithm used for our data to compute a numerical prediction of $s_3$, which serves to account for finite-size effects.  
Our findings are summarized in Fig.~\ref{fig:fig3}, where we present a close-up of a typical measurement near $q_2 = 0$, compared with finite temperature prediction Eq.~\eqref{eq:finiteT_expansion} (solid purple), 2D numerical calculations (dashed light blue), and quasi-2D numerical calculations (dashed red). All predictions use as input the experimentally measured temperature and population in the first $z$-level ($T/T_{\rm F}=0.15(1)$ and $p_1=4.7(4)\%$ for the data shown in the main text)~\cite{supmat}. The first striking observation is that the sole impact of finite temperature is a rounding of the kink within a narrow range of $S_{\rm topo}$ near $q_2=0$.  

This behavior can be understood qualitatively via a Sommerfeld expansion analysis~\cite{supmat}, which shows that when $T\neq0$, there is no power-law temperature correction to the topological formula. This implies that the topology of the Fermi sea, as probed by $s_3$, remains well defined at non-zero temperatures. More quantitatively, $s_3$ can be derived analytically in the special case where $\mathbf{q}_1$ and $\mathbf{q}_2$ form a skinny right triangle with long (short) side of length $|{\bf q}_{1}|$ ($|{\bf q}_{2}|$), respectively, as is the case in Fig.~\ref{fig:fig3}. In the limit $T/T_{\rm F} \ll 2|{\bf q}_1|/k_{\rm F} \ll 1$ and within $S_{\rm topo}$, one obtains 
\beq
s_3({\bf q}_1, {\bf q_2}) = \frac{|{\bf q}_1|| {\bf q}_2|}{(2\pi)^2} 
\coth\!\left( \frac{T_{\rm F}}{T} \frac{|{\bf q}_2|}{k_{\rm F}} \right).
\label{eq:finiteT_expansion}
\eeq
This result shows that the expected topological scaling is recovered away from $q_{2} = 0$, and highlights the absence of any power-law temperature correction to $s_3$, in agreement with the Sommerfeld expansion. Comparison with our strictly 2D numerical calculations shows that finite-size effects further contribute to smoothing the singularity. Finally, comparison of the quasi-2D calculations with our measurements (see Figs.~\ref{fig:fig2} and \ref{fig:fig3}) shows remarkable agreement, providing a convincing explanation for the observed vertical shift as a simple consequence of a small fraction of atoms occupying excited $z$-levels. Strikingly, far from the kink and within $S_{\rm topo}$, all results exhibit the same slope determined by $\chi_{\rm F}=1$.



The excellent agreement with computations for a homogeneous system also indicates that our results are mostly insensitive to the presence of a harmonic trap (IV). This can be qualitatively understood as the measurement of the topological behavior involves wavevectors $\{\textbf{q}_a\}_{a = 1, 2, 3}$ of order $k_{\rm F}$, and thus probes spatial correlations over distances a few $k_{\rm F}^{-1}$ from the central region, where the density variation will be small, leading to the correct value for $s_3$. This reasoning also corroborates the limited impact of temperature and finite-size effects: at finite temperature, the spatial density correlations decay exponentially over the de Broglie wavelength, which for our parameters corresponds to $\sim 10 k_{\rm F}^{-1}$, and we measure correlations in practice for maximum values of $|r_{1,2}|\sim7 k_{\rm F}^{-1}$~\cite{supmat}. Both length scales are comparable, which means correlations are faithfully captured, and both limits will only affect measurements of $s_3$ for small $\{\textbf{q}_a\}_{a = 1, 2, 3}$, as seen in Figure~\ref{fig:fig3}.

Finally, the most surprising observation is the apparent robustness of the Fermi sea topology with respect to inter-spin interactions, as well as the excellent agreement found with the predictions for an ideal gas. As shown in Fig.~\ref{fig:figS5}, a similar level of agreement is observed for ${\cal I} = -0.27(1)$ and $-0.49(2)$, which approach the strongly interacting regime. Recent measurements performed in this regime of interaction revealed that the density-density correlations~\cite{daix2025} and full counting statistics~\cite{dixmerias2025a} within each spin component are almost indistinguishable from those of an ideal Fermi gas. This result, however, lacks a formal theoretical explanation and stands in stark contrast with mean-field predictions based on BCS theory \cite{obeso-jureidini2022}. We believe that the observed robustness of the topological formula Eq.~\eqref{eq:topological_formula} observed here, reinforces and deepens this intriguing connection between properties of the ideal Fermi gas and of a single-spin component of attractive Fermi gases.

\subsection*{Volume Scaling}

An important property of Eq.~(\ref{eq:topological_formula}) and its generalization to $D$ dimensions is that the geometric prefactor in front of $\chi_{\rm F}$ is given by the $D-$volume defined by the triangulating vectors $\{\textbf{q}_p\}_{p = 1, 2, ..., D}$.  To experimentally demonstrate that in our case the geometric factor is set by $\vert \textbf{q}_1 \times \textbf{q}_2\vert=\vert \textbf{q}_1 \vert \vert \textbf{q}_2 \vert  \vert\sin\theta\vert$, we explore a wide range of configurations of the triangulating vectors, by varying the angle $\theta$ between $\textbf{q}_1$ and $\textbf{q}_2$, and measuring the slope of $s_3(q_2)$ as was done previously within the bounds of validity of the topological formula. The results are shown in Fig.~\ref{fig:fig4}, where we report the measured normalized slopes of $(2\pi)^2s_3/{\vert \textbf{q}_1 \vert}$ as function of $\theta$, finding excellent agreement with the predicted result $\vert\sin\theta\vert$ without any fitting parameter. The insets show examples of the rich structure of $s_3$ away from the orthogonal case ($\textbf{q}_1 \perp \textbf{q}_2,\ \theta=\pi/2$). Here again, the $T=0$ prediction Eq.~\eqref{eq:exact_formula_T0} reproduces the qualitative features very well, the quasi-2D numerical calculations show remarkable agreement with the measurements, including the blurring of sharp features and minor shifts, and both predictions display the expected topological behaviour within the bounds of $S_{\rm topo}$. We obtain good agreement with the prediction of Eq.~(\ref{eq:topological_formula}) over a large dynamic range, which is mainly limited by the size of $S_{\rm topo}$ which becomes restrictive for angles far from $\theta = \pi/2\ \rm{mod}\ \pi$. Like before, small systematic shifts of a few percent are expected on the measured slope~\cite{supmat}.

\subsection*{Four-Point Correlations}

We now turn to higher-order correlations. In two dimensions, the Fourier transform of fourth- and higher-order connected correlations measure the Euler characteristic of a torus \cite{tam2024} and are expected to vanish. This result can be understood from the $T=0$ analytical expression \cite{supmat}:
\begin{equation}
s_4(\mathbf{q}_1,\mathbf{q}_2,\mathbf{q}_3) = 
s_3(\mathbf{q}_a,\mathbf{q}_b) + 
s_3(\mathbf{q}_a,\mathbf{q}_c) - 
s_3(\mathbf{q}_a,\mathbf{q}_b + \mathbf{q}_c),
\label{eq:s4}
\end{equation}
where $\mathbf{q}_a$, $\mathbf{q}_b$, and $\mathbf{q}_c$ are appropriately chosen momenta from 
$\{\mathbf{q}_p\}_{p=1,2,3,4}$,
and $\mathbf{q}_4 = -\mathbf{q}_1 - \mathbf{q}_2 - \mathbf{q}_3$, and satisfy $\text{sgn}(\mathbf{q}_a \times \mathbf{q}_b) = \text{sgn}(\mathbf{q}_a \times \mathbf{q}_c) $ . From this expression, it follows straightforwardly that when all $s_3$ terms lie within the topological regime 
and are given by Eq.~\eqref{eq:topological_formula}, their contributions cancel exactly, and the validity criterion of Eq.~\eqref{eq:topological_formula} translates into a distinct validity condition for $s_4=0$~\cite{supmat}.

As was done for the measurement of $s_3$, we directly extract the connected four-point density-density correlation functions from the experimental images and evaluate $s_4(\textbf{q}_1,\textbf{q}_2,\textbf{q}_3)$ via a discrete Fourier transform. Although four-point correlations are challenging to compute experimentally due to the very large space of parameters, we obtain a good signal-to-noise measurement of $s_4$ for a number of configurations (see~\cite{supmat} and Fig.~\ref{fig:figS3}). In Fig.~\ref{fig:fig5}, we show a representative measurement for one configuration of $\{ \textbf{q}_p \}_{p=1, 2, 3, 4}$  where $\textbf{q}_1$ and $\textbf{q}_2$ are fixed and we vary $\textbf{q}_3$ along a line, displaying a clear drop of $s_4$ within the expected validity region (grey area), reaching a value close to zero, with a small offset due to the residual population of atoms in the first excited transverse level. Once again, the $T=0$ prediction features agree very well with our measurements, while the quasi-2D numerical simulations show excellent agreement, and account for the smoothing of sharp features and the vertical offset.

\subsection*{Conclusion and outlook}
We experimentally demonstrate the link between density--density correlation functions and the Fermi sea topology in a quantum gas and obtain the first direct measurement of the Fermi sea's Euler characteristic $\chi_{\rm F}$ in two dimensions. Our results highlight the robustness of this topological relationship with respect to several deviations from ideal conditions, such as sample inhomogeneity and finite temperature. Here, 
topological invariants are obtained from equal-time equilibrium properties of the ideal Fermi gas, as opposed to dynamical quantities, which makes this measurement particularly easy to implement in quantum gas experiments. This connection with topological properties further emphasizes the value of connected correlations in characterizing many-body systems~\cite{schweigler2017,rispoli2019,christakis2023,bureik2025,allemand2025}.

The methods employed here are readily applicable to a wide variety of quantum systems, such as fermions in optical lattices with various geometries and strongly interacting fermions in one, two or three dimensions, and they offer prospects for probing more exotic topologies as well as Lifshitz transitions~\cite{tam2024}. 

Broader questions should be investigated over the longer term. It is remarkable that the prediction of Eq.~(\ref{eq:topological_formula}), established for a translationally invariant system, remains well verified for a harmonically trapped Fermi gas, which raises the question of its validity in the few-fermion regime. Likewise, understanding the impact of interactions on $s_3$, and explaining the observed behavior of same-spin connected correlations -- which remain well described by an ideal Fermi gas even in an interacting regime -- remain open questions. Finally, one could leverage the relation between the Fermi sea topology and multipartite number correlations to measure entanglement in Fermi gases~\cite{tam2022}. 

\subsection*{Acknowledgements} 

We acknowledge Waseem Bakr, Max Prichard and Christophe De Beule for stimulating discussions. T.Y. is grateful to Antoine
Heidmann for his crucial support as head of Laboratoire
Kastler Brossel. This work has been supported by Agence Nationale de la Recherche (Grant No. ANR-21-CE30-0021), CNRS (Tremplin@INP 2020), and R{\'e}gion Ile-de-France in the framework of DIM SIRTEQ and DIM QuanTiP.\\

\textbf{Author contributions:}  C.D., M.D., J.V., performed the experiments. C.D., M.D., J.V., T.d.J., and B.P. all contributed to the analysis of the data. C.D. performed the numerical modelling. P.M.T. and C.L.K. performed theoretical analysis. T.Y. planned and coordinated the study. All authors contributed to the interpretation of the results and to the writing of the manuscript.\\

\textbf{Competing interests:} The authors declare no competing interests.


%


\renewcommand\thefigure{S\arabic{figure}}
\renewcommand\theHfigure{S\arabic{figure}}
\setcounter{figure}{0} 

\renewcommand\theequation{S\arabic{equation}}
\renewcommand\theHequation{S\arabic{equation}}
\setcounter{equation}{0}

\cleardoublepage

\subsection*{Methods}

\subsubsection*{Sample preparation}

The sample preparation was described in \cite{daix2025}. In short, our experiment uses a balanced Fermi mixture of $^6$Li atoms in the first and third lowest-energy hyperfine states, held in a ``light sheet" dipole trap (see Fig.~\ref{fig:fig1}b). We perform evaporative cooling at 690\,G, close to a Feshbach resonance. We subsequently tune the inter-atomic contact interactions via a magnetic field sweep and hold the gas during 500\,ms in order to ensure thermalization. Before imaging the atoms, we selectively remove one spin-state with a resonant light pulse. Toward the end of this blasting pulse, we ramp up the pinning lattice to freeze the motion of the atoms. The pinning lattice is formed by the interference of three lattice arms derived from a 1064\,nm laser which cross each other at 120$^{\circ}$ angles, resulting in a triangular geometry with spacing $a_L = 709$\,nm \cite{jin2024}. The magnetic field is then ramped down to 0\,G in 30\,ms, after which we initiate Raman sideband cooling of the atoms, driving their fluorescence while maintaining them at the bottom of their lattice site. The fluorescence signal is collected on a CCD camera through a microscope objective. The resulting images are analyzed with a high-fidelity neural network to obtain the positions of the individual atoms \cite{verstraten2025}.


\subsubsection*{Computation of $s_3$ and $\chi_{\rm F}$ from experimental data}

We compute three-point connected correlations in real space $\langle N(\textbf{0}) N(\textbf{r}_1) N(\textbf{r}_2) \rangle_{\rm c}$ using the positions of the detected atoms in the pinning lattice. In order to obtain sufficient statistics, the origin coordinate $\textbf{0}$ is taken within a central region of the cloud where the average density is higher than 80\% of the density at the center of the cloud.

We then compute the discrete Fourier transform of these correlations. Owing to the finite size of our cloud, we perform these calculations up to a maximum cutoff distance in real space:

\beq
s_3 = \frac{1}{A_{\rm WS}} 
\sum_{(\textbf{r}_1, \textbf{r}_2) \in \Lambda}
e^{-i\textbf{q}_1\cdot\textbf{r}_1-i\textbf{q}_2\cdot\textbf{r}_2}\langle N(\textbf{0}) N(\textbf{r}_1) N(\textbf{r}_2) \rangle _{\rm c}
\eeq
where $A_{\rm WS}$ is the area of the elementary Wigner-Seitz cell of the pinning lattice and $\Lambda$ is determined by the cutoff condition.

Here, we choose an isotropic cutoff condition $\Lambda = \{(\textbf{r}_1, \textbf{r}_2 )\mid r_1^2+r_2^2 < R_{\rm cutoff}^2 \}$.
The cutoff distance is taken to be $R_{\rm cutoff} \simeq 7 k_{\rm F}^{-1}$, 
or approximately 12 lattice sites. This cutoff distance leads to a finite resolution $\Delta q \simeq 2.75/R_{\rm cutoff}$ in Fourier space. To mitigate finite size effects on the determination of the slope in Fig.~\ref{fig:fig3}, we only fit $s_3$ well within $S_{\rm topo}$, only considering points farther than $\Delta q$ from the edges of $S_{\rm topo}$. We generate a fixed number of points within this subset of $S_{\rm topo}$ and obtain the corresponding slope from a linear fit using ordinary least squares. We have obtained similar results using other cutoff conditions.
In the dilute regime we work in, we can probe wavevectors up to $q_{\rm max}\sim 5 k_{\rm F}$ while fulfilling the Nyquist-Shannon criterion.

To improve our signal-to-noise ratio, we compute $s_3$ for different absolute orientations of $\{\textbf{q}_1,\, \textbf{q}_2\}$ and average the results over several orientations. 
For each configuration $\{\textbf{q}_a\}_{a = 1, 2, 3}$, we compute $s_3$ and an estimate of the associated error via bootstrapping. Sampling from our images, we create 50 batches with 1000 images each and evaluate the resulting connected three-point correlations, from which we obtain $s_3$. This gives us an estimate of the average value and error on $s_3$. We repeat this procedure over 10 different absolute orientations of $\{\textbf{q}_a\}_{a = 1, 2, 3}$. The markers and errorbars shown for $s_3$ correspond to the resulting mean and standard error of $s_3$, computed using inverse variance weighting.

\subsubsection*{Computation of $s_4$}

$s_4$ is computed similarly to $s_3$. We start by evaluating four-point connected correlations in real space $\langle N(\textbf{0}) N(\textbf{r}_1) N(\textbf{r}_2) N(\textbf{r}_3)\rangle_{\rm c}$. The origin coordinate $\textbf{0}$ is taken within the same central region as for $s_3$. We then compute the discrete Fourier transform:
\beqa
s_4 = \frac{1}{A_{\rm WS}} 
\sum_{(\textbf{r}_1, \textbf{r}_2, \textbf{r}_3) \in \Lambda}
&\Big( \langle N(\textbf{0}) N(\textbf{r}_1) N(\textbf{r}_2) N(\textbf{r}_3) \rangle _{\rm c} \nonumber\\
&\times\,  e^{-i\textbf{q}_1\cdot\textbf{r}_1-i\textbf{q}_2\cdot\textbf{r}_2-i\textbf{q}_3\cdot\textbf{r}_3} \Big)
\eeqa
with $\Lambda = \{(\textbf{r}_1, \textbf{r}_2, \textbf{r}_3 )\mid r_1^2+r_2^2+r_3^2 < R_{\rm cutoff}^2 \}$. Similar to $s_3$, the cutoff distance is $R_{\rm cutoff} \simeq 7 k_{\rm F}^{-1}$, and we average $s_4$ over 10 absolute orientations of $\{\textbf{q}_p\}_{p = 1, 2, 3,4}$ ($\textbf{q}_4=-\textbf{q}_1-\textbf{q}_2-\textbf{q}_3$). The errorbars shown correspond to the standard error of the mean over those orientations. Additional examples of $s_4$ measurements are shown in Fig.~\ref{fig:figS3}.

\subsubsection*{Numerical calculations}

Using Wick's theorem, $\langle n(\textbf{0}) n(\textbf{r}_1) n(\textbf{r}_2) \rangle_{\rm c}$ and $\langle n(\textbf{0}) n(\textbf{r}_1) n(\textbf{r}_2) n(\textbf{r}_3) \rangle_{\rm c}$ can be expressed in terms of the field correlator \mbox{$g_1(\textbf{r}'-\textbf{r}) = \langle \psi ^{\dagger}(\textbf{r}') \psi (\textbf{r}) \rangle$}, which can be efficiently computed for a 2D ideal Fermi gas at finite temperature. We compute the theoretical value of connected correlations for a homogeneous system with our experimental parameters (average density, reduced temperature) and sampled on our pinning lattice. We then apply the same Fourier transform algorithm used for the experimental data. In this procedure, finite-size and sampling effects on the discrete Fourier transform are included by design.

In this framework, the effect of populated excited transverse levels can be evaluated. For uncorrelated layers, the total connected density-density correlations are the sum of the connected density-density correlations for each layer. For the datasets presented here, the proportion of atoms in $n_z = 0$ is $\gtrsim 95\%$, as obtained from a two-parameter fit of $\langle N(\textbf{0}) N(\textbf{0}) N(\textbf{r}) \rangle_{\rm c}$ which we also use to obtain an estimate of the average reduced temperature $T/T_{\rm F}$. For the wavevectors probed in this work ($|\textbf{q}_1| \geq 0.5\,k_{\rm F}$), the topological condition Eq.~(\ref{eq:validity_criterion}) is not fulfilled for the smaller Fermi seas associated with higher transverse motional levels. As a consequence, their contribution to $s_3$ is almost flat and mainly leads to an offset of $s_3$, while marginally affecting its slope, especially for larger values of $|\textbf{q}_1|$. By fitting the obtained numerical results in the same way as the data, we obtain an estimate of systematic shifts $\Delta\chi_{\rm F}$ on the determination of $\chi_{\rm F}$, which are on the order of a few percent. The mean shifts are reported with their standard deviation in Table~\ref{tab:table1}.

\begin{table}[h!]
    \begin{tabular}{|c|c|c|c|c|c|}
        \hline
        $|\textbf{q}_1|/k_{\rm F}$ & 0.75 & 1.00 & 1.25 & 1.50 & 1.75 \\
        \hline
        $\Delta\chi_{\rm F}$ & +0.02(3) & +0.01(3) & 0.00(5) & 0.00(4) & -0.03(3) \\
        \hline
    \end{tabular}
    \label{tab:table1}
    \caption{\textbf{Systematic shifts on the determination of $\chi_{\rm F}$.} Value of $\Delta\chi_{\rm F}$ obtained from a linear fit to our quasi-2D numerical results for each value of $|\textbf{q}_1|/k_{\rm F}$ shown in Fig.~\ref{fig:fig4}. The effect of atoms in excited transverse levels together with finite size effects lead to small systematic deviations that depend on the exact configuration of $\{\textbf{q}_a\}_{a = 1, 2, 3}$. While the values reported are consistent with $\chi_{\rm F}=1$ within errorbars, we observe that small values of $|\textbf{q}_1|/k_{\rm F}$ lead to an overestimation of $\chi_{\rm F}$, with the opposite trend for values of $|\textbf{q}_1|/k_{\rm F}$ close to $2$.}
\end{table}

\subsubsection*{Exact formula for $s_3$ at T = 0}

In this section we derive the compact analytic formula for three-point correlations $s_3({\bf q}_1,{\bf q}_2)$ at the zero temperature, for generic configurations of $\{\textbf{q}_a\}_{a = 1, 2, 3}$. The general formula for the three point correlation is \cite{tam2022, tam2024}
\begin{equation}\begin{split}
    s_3({\bf q}_1,{\bf q}_2) = \int \frac{d^2{\bf k}}{(2\pi)^2} 
    (&f_0 - f_0 f_1 - f_0 f_2 - f_0 f_{12}\\
   & + f_0 f_1 f_{12} + f_0 f_2 f_{12})
    \label{s3formula}
\end{split}\end{equation}
where $f_{a} \equiv f_{{\bf k}+{\bf q}_a }\equiv f(E_{{\bf k}+{\bf q}_a})$ is the Fermi occupation factor, and $f_0 \equiv f_{\bf k}$. At zero temperature, $f_a = \Theta(E_{\rm F}-E_{{\bf k}+{\bf q}_a})$ is a Heaviside step function.  

Provided $k_{\rm F}$ is large enough (or $\{\textbf{q}_a\}_{a = 1, 2, 3}$ are small enough), it was shown in Ref. \cite{tam2024} that at zero temperature this is given by the topological formula:
\begin{equation}
    s_3({\bf q}_1,{\bf q}_2) = \frac{|{\bf q}_1\times {\bf q}_2|}{(2\pi)^2}.
    \label{s3 topo}
\end{equation}
For a circular Fermi surface, the precise condition for the topological formula to be valid is (with the corresponding region in momentum space denoted as $S_{\rm topo}$)
\begin{equation}\label{eq: topological condition}
    k_{\rm F} > R_{\{{\bf q}\}} \equiv \frac{|{\bf q}_1||{\bf q}_2||{\bf q}_3|}{2|{\bf q}_1 \times {\bf q}_2|},
\end{equation}
where ${\bf q}_3 = - {\bf q}_1 - {\bf q}_2$.   $R_{\{{\bf q}\}}$ is the radius of the circle that circumscribes the triangle formed by $\{\textbf{q}_a\}_{a = 1, 2, 3}$. When Eq. \eqref{eq: topological condition} is satisfied (i.e., inside $S_{\rm topo}$), the mesh for the triangulation argument given in Ref. \cite{tam2024} is fine enough for Eq. \eqref{s3 topo} to be valid. Next we derive a more general analytic expression for $s_3$ that is valid even outside $S_{\rm topo}$.

At zero temperature for a circular Fermi surface, Eq. \eqref{s3formula} involves the area of the intersection of circles:
\begin{equation}\begin{split}
    s_3({\bf q}_1,{\bf q}_2) = \frac{1}{(2\pi)^2}\Big(&A_1 - A_2({\bf q}_1) - A_2({\bf q}_2)\\
    & - A_2({\bf q}_{12}) + 2 A_3({\bf q}_1,{\bf q}_2) \Big),
    \label{area sum}
\end{split}\end{equation}
where $A_n({\bf q}_a)$ is the intersection area of $n$ circles separated by vectors ${\bf q}_a$, and we note that $A_3({\bf q}_1,{\bf q}_2) = A_3({\bf q}_2,{\bf q}_1)$.

\begin{figure}
    \centering
    \includegraphics[width=0.9\linewidth]{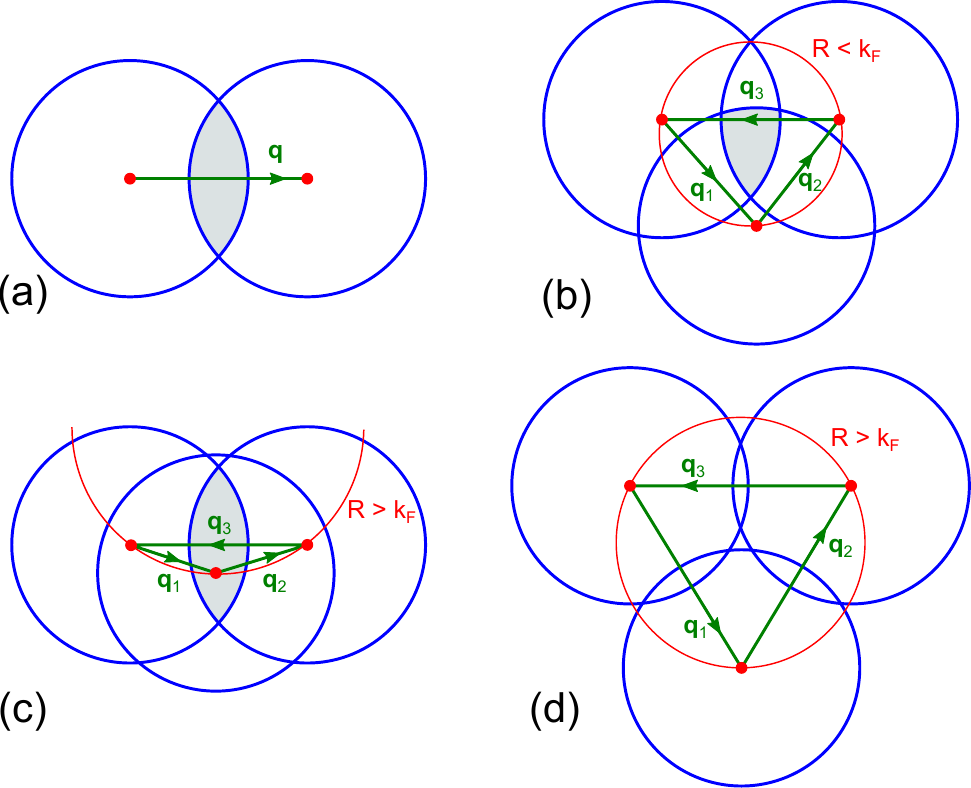}
    \caption{Intersection area (labeled as the shaded region) of two circles (a) and three circles for three configurations (b,c,d). }
    \label{fig:2and3circles}
\end{figure}

Clearly, $A_1 = \pi k_{\rm F}^2$.   $A_2$ (see Fig. \ref{fig:2and3circles}(a)) can be found using elementary geometry:
\begin{equation}
    A_2({\bf q}) =  \pi k_{\rm F}^2 G\left(\frac{|{\bf q}|}{2k_{\rm F}}\right) 
\end{equation}
with
\begin{equation}
    G(X) =  \frac{2}{\pi}\left(\cos^{-1} X - X\sqrt{1-X^2}\right)\Theta(1-X)
\end{equation}
where $\Theta(1-X)$ is a step function.

$A_3$ is the intersection area of three circles, which is a harder geometry problem when all three circles are involved (see Fig. \ref{fig:2and3circles}(b)).  But this occurs precisely when  $R_{\{{\bf q}\}}<k_{\rm F}$, where the topological formula already gives the answer.  Therefore,
\begin{equation}\begin{split}
    A_3({\bf q}_1,{\bf q}_2) = \frac{1}{2}\Big(& |{\bf q}_1 \times {\bf q}_2| + A_2({\bf q}_1) + A_2({\bf q}_2)\\
    &+ A_2({\bf q}_3) - A_1\Big).
\end{split}\end{equation}
One can also solve the geometry problem directly and obtain the same result.

For $R_{\{{\bf q}\}}>k_{\rm F}$, $A_3$ reduces to the intersection area of two circles.  The result depends on whether the triangle formed by $\{\textbf{q}_a\}_{a = 1, 2, 3}$ is obtuse (has an angle $> \pi/2$, see Fig. \ref{fig:2and3circles}(c)) or acute (all angles $< \pi/2$, see Fig. \ref{fig:2and3circles}(d)).  
If we assume (without loss of generality) that $|{\bf q}_3|$ is the maximum of $|{\bf q}_{1,2,3}|$, then,
\begin{equation}
    A_3({\bf q}_1,{\bf q}_2) =\left\{\begin{array}{cc}
        A_2({\bf q}_3) & {\bf q}_1\cdot{\bf q}_2  >0 \\
        0 & {\bf q}_1\cdot{\bf q}_2  <0
    \end{array} \right.
\end{equation}
It follows from (\ref{area sum}) that for $R_{\{{\bf q}\}}> k_{\rm F}$
\begin{equation}\begin{split}
    s_3({\bf q}_1,{\bf q}_2) = \frac{1}{(2\pi)^2}\Big(&A_1 - A_2({\bf q}_1)  - A_2({\bf q}_2) \\
  & + {\rm sgn}[{\bf q}_1\cdot{\bf q}_2] A_2({\bf q}_3)\Big).
\end{split}\end{equation}
This can be written symmetrically using the fact that ${\bf q}_1\cdot {\bf q}_3<0$ and ${\bf q}_2\cdot{\bf q}_3<0$, given the above assumption that $|{\bf q}_3|$ is the maximum of $|{\bf q}_{1,2,3}|$.   If we define $\sigma_{a} = {\rm sgn}[{\bf q}_b\cdot{\bf q}_c]$ for $a\ne b\ne c = 1,2,3$, the general case for
$R_{\{{\bf q}\}}> k_{\rm F}$ can be written as
\begin{equation}
    s_3({\bf q}_1,{\bf q}_2) = 
    \frac{k_{\rm F}^2}{4\pi}\left(1 + \sum_{a=1}^3 \sigma_a G\left(\frac{|{\bf q}_a|}{2k_{\rm F}}\right) \right).
    \label{s3 formula}
\end{equation}
Combined with Eq. \eqref{s3 topo}, we obtain Eq. \eqref{eq:exact_formula_T0} in the main text. 

\subsubsection*{Sommerfeld expansion}
In this section we establish the robustness of $s_3({\bf q}_1,{\bf q}_2)$ for probing the Fermi sea topology at a finite temperature, via the Sommerfeld expansion, showing that there is no power-law in $T$ correction (in the topological regime $S_{\rm topo}$ satisfying Eq. \eqref{eq: topological condition}).
The Sommerfeld expansion amounts to the following expansion for the finite-temperature Fermi-Dirac distribution: $f(E,T)=f^{(0)}(E)+\delta f(E,T)$, 
\begin{equation}\label{eq: Sommerfeld of FD}
    \delta f(E,T)= - \sum_{n=1}^\infty b_{2n}T^{2n}\delta^{(2n-1)}(E-E_{\rm F}),
\end{equation}
where $f^{(0)}(E) = \Theta(E_{\rm F}-E)$ is the zero-temperature distribution, $\delta f$ captures all the power-law finite-temperature corrections, with coefficients $b_2=\pi^2/(6k_B^2)$, $b_4=7\pi^4/(360k_B^4)$, etc. We adopt a simplified notation $\delta^{(2n-1)}(E-E_{\rm F}) \equiv (\frac{d}{dE})^{2n-1} \delta(E-E_{\rm F})$. Importantly, the power-law in $T$ corrections arise only at $E=E_{\rm F}$ (i.e., the Fermi surface), which is an essential property we are going to invoke below. 

Plugging in the Sommerfeld expansion into the general expression for $s_3$ in Eq. \eqref{s3formula}, we obtain
\begin{widetext}
\begin{equation}\label{eq: s3 at finite T}
\begin{alignedat}{2}
    &s_3({\bf q}_1,{\bf q}_2) = \int \frac{d^2{\bf k}}{(2\pi)^2}(f^{(0)}_{\bf k}+\delta f_{\bf k})(1-f^{(0)}_{{\bf k}+{\bf q}_{12}}-\delta f_{{\bf k}+{\bf q}_{12}})(1-f^{(0)}_{{\bf k}+{\bf q}_1}-f^{(0)}_{{\bf k}+{\bf q}_2}-\delta f_{{\bf k}+{\bf q}_1}-\delta f_{{\bf k}+{\bf q}_2})  \\
    &= s^{(0)}_3({\bf q}_1,{\bf q}_2) + \int \frac{d^2{\bf k}}{(2\pi)^2} \delta f_{\bf k} (1-f^{(0)}_{{\bf k}+{\bf q}_{12}}-f^{(0)}_{{\bf k}+{\bf q}_{1}}-f^{(0)}_{{\bf k}+{\bf q}_{2}}-f^{(0)}_{{\bf k}-{\bf q}_{12}}-f^{(0)}_{{\bf k}-{\bf q}_{1}}-f^{(0)}_{{\bf k}-{\bf q}_{2}}\\
    &\quad\quad\quad\quad\quad\quad\quad\quad\quad\quad+f^{(0)}_{{\bf k}+{\bf q}_{1}}f^{(0)}_{{\bf k}+{\bf q}_{12}}+f^{(0)}_{{\bf k}+{\bf q}_{2}}f^{(0)}_{{\bf k}+{\bf q}_{12}}+f^{(0)}_{{\bf k}+{\bf q}_{2}}f^{(0)}_{{\bf k}-{\bf q}_{1}}+f^{(0)}_{{\bf k}-{\bf q}_{2}}f^{(0)}_{{\bf k}+{\bf q}_{1}}+f^{(0)}_{{\bf k}-{\bf q}_{12}}f^{(0)}_{{\bf k}-{\bf q}_{1}}+f^{(0)}_{{\bf k}-{\bf q}_{12}}f^{(0)}_{{\bf k}-{\bf q}_{2}})\\
    &- \int \frac{d^2{\bf k}}{(2\pi)^2} \left[\delta f_{\bf k} \delta f_{{\bf k}+{\bf q}_1} (1-f^{(0)}_{{\bf k}+{\bf q}_{12}}-f^{(0)}_{{\bf k}-{\bf q}_{2}}) +\delta f_{\bf k} \delta f_{{\bf k}+{\bf q}_2} (1-f^{(0)}_{{\bf k}+{\bf q}_{12}}-f^{(0)}_{{\bf k}-{\bf q}_{1}}) +\delta f_{\bf k} \delta f_{{\bf k}+{\bf q}_{12}} (1-f^{(0)}_{{\bf k}+{\bf q}_{1}}-f^{(0)}_{{\bf k}+{\bf q}_{2}})\right]
\end{alignedat}
\end{equation}
\end{widetext}
where $s^{(0)}_3$ is the zero-temperature three-point correlation, and we used the shorthand notation ${\bf{q}}_{12}={\bf{q}}_1+{\bf{q}}_2$. Notice that, in the above, we have only kept terms with up to two $\delta f$'s. This is because, according to Eq. \eqref{eq: Sommerfeld of FD}, terms with three $\delta f$'s, $\delta f_{\bf k} \delta f_{{\bf k}+{\bf q}_a}\delta f _{{\bf k}+{\bf q}_{12}}$ ($a=1,2$), would contribute only when the three points ${\bf k}, {\bf k}+{\bf q}_{a}, {\bf k}+{\bf q}_{12}$ are simultaneously situated on the $T=0$ Fermi surface, i.e., when the triangle formed by ${\bf q}_{1}, {\bf q}_{2}$ and ${\bf q}_{3}$ is being inscribed by the Fermi surface. But this is precisely the critical condition for the validity of our topological formula, see Eq. \eqref{eq: topological condition}. We assume the Fermi surface is large enough so that we are within the topological regime at $T=0$, and this implies the product of three $\delta f$'s is always zero.

\begin{figure}[H]
    \centering
    \resizebox{\columnwidth}{!}{\includegraphics[]{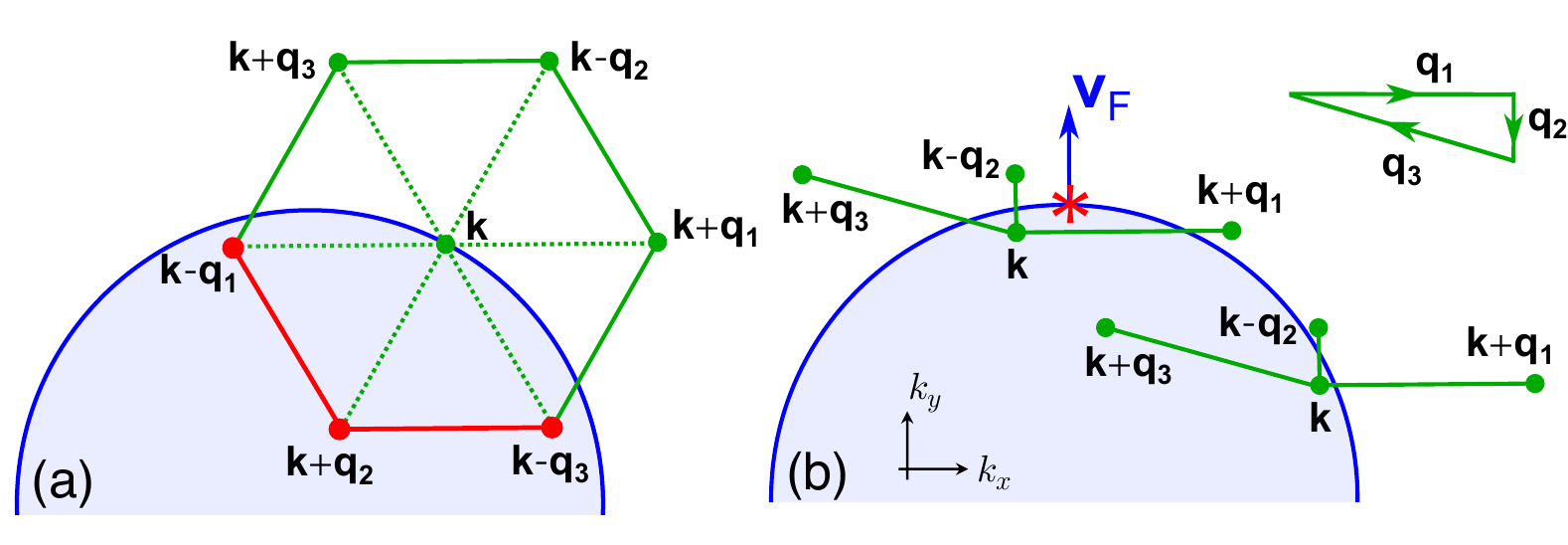}}
    \caption{(a) Neighbors of ${\bf k}$ form a 1D simplicial complex consisting of vertices and solid edges. When ${\bf k}$ lies on the Fermi surface, we are guaranteed to have $\chi_\text{neighbors of ${\bf k}$}=1$ for the part of the complex (labeled red) lying inside the Fermi sea (as the blue shaded region). (b) When ${{\bf q}_1, {\bf q}_2, {\bf q}_{3}=-{\bf q}_{1}-{\bf q}_{2}}$ form a skinny right-angle triangle ($|{\bf q}_2| \ll |{\bf q}_{1,3}|$) as depicted on the upper right corner, the integral in $s_3(\{{\bf q}\})$ is dominated by the neighborhood Fermi surface critical points (labeled as the red asterisk)  with ${\bf v}_{\bf k} \cdot {\bf q}_{1}= 0$. Away from the critical points, the integrand which is proportional to $1-f_{{\bf k}+{\bf q}_1}-f_{{\bf k}+{\bf q}_3}$ is exponentially suppressed as $e^{-\beta v_{\rm F} |{\bf q}_1|}$.}
    \label{FigS3}
\end{figure}

Now consider the $\delta f$-terms in Eq. \eqref{eq: s3 at finite T}: the integrand can be recognized as $    \delta f_{\bf k} \cdot (1-\chi_{\text{neighbors of {\bf k}}})$, where
\begin{equation}
\begin{split}
\chi_{\text{neighbors of {\bf k}}} = \left(\sum_{a=1}^3f^{(0)}_{{\bf k}+{\bf q}_a} + \sum_{a=1}^3f^{(0)}_{{\bf k}-{\bf q}_a}\right)\\ - \left( \sum_{a\neq b =1}^3 f^{(0)}_{{\bf k}+{\bf q}_a}f^{(0)}_{{\bf k}-{\bf q}_b}\right)
\end{split}
\end{equation}
is simply the Euler characteristic of the 1D simplicial complex (a set of vertices and edges) that is connected to ${\bf k}$ by $\{\textbf{q}_a\}_{a = 1, 2, 3}$ and lying within the Fermi sea. When $\{\textbf{q}_a\}_{a = 1, 2, 3}$ are small enough (or the Fermi surface is large enough) such that the zero-temperature $s^{(0)}_3$ is topological, i.e. satisfying Eq. \eqref{eq: topological condition}, we always have $\chi_{\text{neighbors of {\bf k}}}=1$. One scenario is illustrated in Fig. \ref{FigS3}(a).  Hence the $\delta f$-terms in Eq. \eqref{eq: s3 at finite T} are exactly canceled out.

Finally, we consider the $\delta f_{{\bf k}}\;\delta f_{{\bf k}+{\bf q}_a}$-terms in Eq. \eqref{eq: s3 at finite T}. For concreteness, first take $a=1$. The delta-functions in $\delta f$ dictates that both ${\bf k}$ and ${\bf k}+{\bf q}_{1}$ are on the Fermi surface. When the topological condition is satisfied, then one of the points $\{{\bf k}-{\bf q}_2, {\bf k}+{\bf q}_{12}\}$ must be lying inside the Fermi sea while another must be lying outside, hence $1-f^{(0)}_{{\bf k}+{\bf q}_{12}}-f^{(0)}_{{\bf k}-{\bf q}_{2}} = 0$. The same argument applies to other $\delta f_{{\bf k}}\;\delta f_{{\bf k}+{\bf q}_a}$-terms. 

All in all, we conclude that all power-law corrections from the Sommerfeld expansion are completely canceled out in Eq. \eqref{eq: s3 at finite T}, given that we are in the topological regime where $s_3^{(0)}({\bf q}_1,{\bf q}_2) =|{\bf q}_1 \times {\bf q}_2|/(2\pi)^2$.   Note that outside the topological regime specified by Eq. \eqref{eq: topological condition}, the three points of the triangle defined by $\{\textbf{q}_a\}_{a = 1, 2, 3}$ can be at the Fermi surface, so there will be power law in $T$ corrections to $s_3({\bf q}_1,{\bf q}_2)$. 

Inside the topological regime there will be exponentially small corrections, of order $e^{-T_0(\{{\bf q}\})/T}$, where $T_0(\{{\bf q}\})\rightarrow 0$ at the boundary of the topological regime. Deep inside the topological regime, with $R_{\{{\bf q}\}} \ll k_{\rm F}$, it can be argued that the energy scale for the exponential suppression (which is associated to the product of three $\delta f$'s) is given by $T_0 \sim \hbar v_{\rm F} q_\perp$, where $q_\perp$ is the minimal height of the triangle formed by $\{\textbf{q}_a\}_{a = 1, 2, 3}$. In the following section we derive an explicit form for the exponential suppression of thermal corrections for a specific class of  configurations of $\{\textbf{q}_a\}_{a = 1, 2, 3}$ inside the topological regime.

\subsubsection*{Analytic formula for $s_3$ at finite temperature}

We consider the case where ${\bf q}_1 \perp {\bf q}_2$ and $\{\textbf{q}_a\}_{a = 1, 2, 3}$ form a skinny right-angle triangle as depicted in Fig. \ref{FigS3}(b), with $|{\bf q}_2| \ll |{\bf q}_{1,3}|$. For $|{\bf q}_{1,3}| \ll k_{\rm F}$, this geometry is then deep inside the topological regime, which we shall focus on. Let us rewrite Eq. \eqref{s3formula} as follows
\begin{equation}\label{eq: s3_permuted}
\begin{split}
    s_3({\bf q}_1,{\bf q}_2) &= \int \frac{d^2{\bf k}}{(2\pi)^2} f_{\bf k} (1-f_{{\bf k}-{\bf q}_{3}})(1-f_{{\bf k}+{\bf q}_1}-f_{{\bf k}+{\bf q}_2})\\
    &=\int \frac{d^2{\bf k}}{(2\pi)^2} f_{\bf k} (1-f_{{\bf k}-{\bf q}_{2}})(1-f_{{\bf k}+{\bf q}_1}-f_{{\bf k}+{\bf q}_3}),
\end{split}
\end{equation}
with $f_{\bf k} = [e^{\beta (E_{\bf k}-E_{\rm F})}+1]^{-1}$ the Fermi occupation at inverse temperature $\beta=1/(k_B T)$. In the second equality we have used the property that $s_3({\bf q}_1,{\bf q}_2)$ is invariant under permutations of $\{\textbf{q}_a\}_{a = 1, 2, 3}$. Using this form, it can be appreciated in Fig. \ref{FigS3}(b) that the integral in $s_3({\bf q}_1,{\bf q}_2)$ is dominated by Fermi surface critical points where ${\bf v}_{\bf k} \cdot {\bf q}_1=0$. This is because the factor $f_{\bf k} (1-f_{{\bf k}-{\bf q}_{2}})$ in the integrand restricts ${\bf k}$ to be near the Fermi surface, and if ${\bf k}$ is far from such critical points then either ${\bf k}+{\bf q}_1$ is inside the Fermi sea (and ${\bf k}+{\bf q}_3$ is outside) or ${\bf k}+{\bf q}_3$ is inside the Fermi sea (and ${\bf k}+{\bf q}_1$ is outside), leading to a suppression of $1-f_{{\bf k}+{\bf q}_1}-f_{{\bf k}+{\bf q}_3}$ in both cases. 

Assuming $\beta v_{\rm F} |{\bf q}_1| \gg 1$, we can thus focus on the neighborhood of one such critical point (i.e., the red asterisk in Fig. \ref{FigS3}(b)) and linearize the dispersion there as (take $\hbar=1$)
\begin{equation}
\begin{split}
    E_{{\bf k}+{\bf q}}-E_{\rm F} &=v_{\rm F}(|{\bf k}+{\bf q}|-k_{\rm F}) \\
    &\approx v_{\rm F}(k_y-k_{\rm F}+q_y+\frac{(k_x+q_x)^2}{2k_{\rm F}})
\end{split}
\end{equation}
The coordinate system is set up such that ${\bf q}_{1} \parallel \hat{k}_{x}$ and ${\bf q}_{2} \parallel \hat{k}_{y}$. We introduce the following variables for the sake of simplifying finite-temperature expressions:
\begin{equation}
\begin{split}
    y&\equiv e^{\beta v_{\rm F}(k_y-k_{\rm F})+\frac{k_x^2}{2k_{\rm F}}},\;\; x_a\equiv e^{\frac{\beta v_{\rm F}q_{a,x}}{k_{\rm F}}k_x}, \\
    \delta_a &\equiv e^{\beta v_{\rm F}q_{a,y}},\;\;\;\;\;\;\;\;\;\;\;\;\;\;\; \gamma_a \equiv e^{\beta v_{\rm F} \frac{q_{a,x}^2}{2k_{\rm F}}},
\end{split}
\end{equation}
for $a=1,2,3$. Furthermore, define $\alpha_a \equiv \delta_a \gamma_a x_a$. The integral in Eq. \eqref{eq: s3_permuted} can then be written as (notice that $\delta_3=1/\delta_2$)
\begin{equation}
\begin{split}
    &s_3({\bf q}_1,{\bf q}_2) = \frac{1}{(2\pi)^2}\frac{1}{\beta v_{\rm F}}\int dk_x \\
    &\int_0^\infty dy \frac{\delta_{3}}{(y+1)(\delta_{3}y+1)}\left[1-\frac{1}{\alpha_1 y +1}-\frac{1}{\alpha_3 y +1}\right].
\end{split}
\end{equation}
First of all, the $y$-integral can be performed analytically to give
\begin{widetext}
\begin{equation}
\begin{split}
    &-\frac{\gamma _1^2 x_1^2 \log \left(\frac{\gamma _1 x_1}{\delta
   _3}\right)}{\left(\gamma _1 x_1-1\right) \left(\gamma _1 x_1-\delta
   _3\right)}+\frac{x_1^2 \log \left(x_1\right)}{\left(x_1-\gamma _1\right)
   \left(x_1-\gamma _1 \delta _3\right)}-\frac{x_1^2 \log \left(\gamma _1 \delta
   _3\right)}{\left(x_1-\gamma _1\right) \left(x_1-\gamma _1 \delta
   _3\right)}\\
   &-\frac{\gamma _1^2 x_1 \log \left(\delta _3\right)}{\left(\delta
   _3-1\right) \left(x_1-\gamma _1\right) \left(\gamma _1 x_1-1\right)}+\frac{x_1 \log
   \left(\delta _3\right)}{\left(\delta _3-1\right) \left(x_1-\gamma _1\right)
   \left(\gamma _1 x_1-1\right)}-\frac{x_1 \log \left(x_1\right)}{x_1-\gamma
   _1}+\frac{x_1 \log \left(\gamma _1\right)}{x_1-\gamma _1}+\frac{\gamma _1 x_1 \log
   \left(\gamma _1 x_1\right)}{\gamma _1 x_1-1},
\end{split}
\end{equation}
\end{widetext}
where simplifications have been attained using the right-angle triangle geometry (i.e., $\gamma_3=\gamma_1$, $x_3=1/x_1$, $\delta_1=1$). Next, the $k_x$-integral can be performed via the change of variable: 
$\int_{-\infty}^\infty dk_x = \frac{k_{\rm F}}{\beta v_{\rm F}|{\bf q}_{1}|}\int_{0}^\infty dx_1 \frac{1}{x_1}$.  We thus arrive at
\begin{equation}
\begin{split}
    s_3({\bf q}_1,{\bf q}_2) &= \frac{1}{(2\pi)^2} \frac{1}{\beta v_{\rm F}}\frac{k_{\rm F}}{\beta v_{\rm F} |{\bf q}_{1}|}\cdot \frac{2 \log\gamma_1 \log \delta_3}{1-\delta_3^{-1}} \\
    &= \frac{|{\bf q}_{1}| q_{2}}{(2\pi)^2} \frac{1}{e^{\beta v_{\rm F} q_{2}}-1},
\end{split}
\end{equation}
with $q_2=q_{2,y}$.
Notice that the above term is just the contribution from one critical point with ${\bf v}_{\rm F} \parallel \hat{y}$. For circular Fermi surfaces of interest in this work, we should also account for the contribution from the other critical point with ${\bf v}_{\rm F} \parallel -\hat{y}$, which can be obtained by flipping $q_{2}\rightarrow -q_{2}$ in the above analysis. All things considered, we arrive at (with $\hbar$ restored)
\begin{equation}
    s_3({\bf q}_1,{\bf q}_2) = \frac{|{\bf q}_{1}| |{\bf q}_{2}|}{(2\pi)^2} \coth{\frac{\beta \hbar v_{\rm F}|{\bf q}_{2}|}{2}}.
\end{equation}
Finally, with $\hbar v_{\rm F} = 2 k_B T_{\rm F} /k_{\rm F}$, we obtain Eq. \eqref{eq:finiteT_expansion} in the main text. For very low $T$ ($ \beta \hbar v_{\rm F}|{\bf q}_2| \gg 1$), this modifies the topological formula by an exponentially suppressed thermal correction, $\exp\left(-\frac{2T_{\rm F}}{T}\frac{|{\bf q}_2|}{k_{\rm F}}\right)$, consistent with our Sommerfeld expansion analysis presented in the previous section. Equation  \eqref{eq:finiteT_expansion} is more general in that it does not make any assumption about $\beta \hbar v_{\rm F} |{\bf q}_2|$, it is valid as along as $\beta \hbar v_{\rm F} |{\bf q}_1| \gg 1$. 

\subsubsection*{Exact formula for $s_4$ at T = 0}

In Ref. \cite{tam2024} it was argued that in two dimensions the four point correlation $s_4({\bf q}_1,{\bf q}_2,{\bf q}_3)$ vanishes, provided $\{\textbf{q}_p\}_{p = 1, 2, 3,4}$, with $\textbf{q}_4=-\sum_{i=1}^3 \textbf{q}_i$ are small enough to be in the topological regime.   Here we provide a general expression for $s_4$ that is valid even outside the topological regime, and defines the precise boundaries of the topological regime.   We first state the result, and then outline the derivation.\\

\textbf{Result:} The four-point equal time correlation can be expressed in terms of the three-point correlator, $s_3$, given in Eq. \eqref{s3 topo} or Eq. \eqref{s3 formula}.  We find
\begin{equation}
    s_4({\bf q}_1,{\bf q}_2, {\bf q}_3) = s_3({\bf q}_a,{\bf q}_b) + s_3({\bf q}_a,{\bf q}_c) - s_3({\bf q}_a,{\bf q}_b+{\bf q}_c)
    \label{s4 formula}
\end{equation}
where ${\bf q}_a$, ${\bf q}_b$ and ${\bf q}_c$ are chosen from ${\bf q}_1$, ${\bf q}_2$, ${\bf q}_3$ and ${\bf q}_4 = -{\bf q}_1-{\bf q}_2 - {\bf q}_3$ according to the following algorithm:

\begin{itemize}

\item[1.]   Construct the matrix 
\begin{equation}
    C_{ij} \equiv C({\bf q}_i,{\bf q}_j) \equiv \frac{2k_{\rm F} ({\bf q}_i \times {\bf q}_j )}{|{\bf q}_i| |{\bf q}_j| |{\bf q}_i+{\bf q}_j|}.
    \label{cij}
\end{equation}   
This defines an antisymmetric $4 \times 4$ matrix that depends on 6 numbers.  $ k_{\rm F}/|C_{ij}| $ is equal to the radius $R({\bf q}_i,{\bf q}_j)$ of the circle that circumscribes the triangle with sides ${\bf q}_i$, ${\bf q}_j$, $-{\bf q}_i-{\bf q}_j$.    Thus, the topological condition for $s_3({\bf q}_i,{\bf q}_j)$ is that $|C_{ij}| > 1$.       These 6 numbers are the data needed to determine  $a$, $b$ and $c$.

\item[2.]  Determine the {\it largest} element $C_>$ of $C_{ij}$  (by definition positive), along with the {\it next largest} element $C'_>$ of $C_{ij}$ (also positive) that is in the same row or column as $C_>$.     

\item[3.]  $a$ is the position of the row or column that contains $C_>$ and $C'_>$.     

\item[4.]  $b$ is the position of $C_>$ in the $a$’th row (column).

\item[5.]  $c$ is the position of $C'_>$ in the $a$’th row (column).
\end{itemize}

It can be seen that if all three $s_3$’s in Eq. \eqref{s4 formula} are in the topological regime, then everything cancels, and $s_4=0$.     Thus the ‘topological condition’ for $s_4$ is:
\begin{equation}
 |C({\bf q}_a,{\bf q}_b)|>1, \  |C({\bf q}_a,{\bf q}_c)| >1,  \ |C({\bf q}_a,{\bf q}_b+{\bf q}_c)|>1.
\end{equation}
The last inequality can also be written as $|C({\bf q}_a,{\bf q}_d)| > 1$, where ${\bf q}_d = -{\bf q}_a-{\bf q}_b-{\bf q}_c$.   Thus, the topological condition can be stated as:

\begin{quote}
 $s_4({\bf q}_1,{\bf q}_2,{\bf q}_3) = 0$ if all the non-zero elements of the row (column) of $C_{ij}$ that contains $C_>$ and $C'_>$ satisfy $|C_{ij}| > 1$.
\end{quote}
The above serve as the main results on $s_4$ that we have highlighted in the main text around Eq. \eqref{eq:s4}, and the derivation is outlined below.\\ 

\textbf{Derivation:} Using the short-hand notation that $f_{a} \equiv f_{{\bf k}+{\bf q}_a }\equiv f(E_{{\bf k}+{\bf q}_a})$ is the Fermi occupation factor, and $f_0 \equiv f_{\bf k}$, the general formula for $s_4({\bf q}_1,{\bf q}_2,{\bf q}_3)$ is,
\begin{widetext}
\begin{equation}
\begin{split}
     s_4 = \int \frac{d^2{\bf k}}{(2\pi)^2}  f_0 \Big(& 1 - (f_1 + f_2+ f_3 + f_{12} + f_{13} + f_{23} + f_{123}) \\
      & + f_1 ( f_{12} +f_{13} + f_{123} )+ f_2 (f_{12} + f_{23} + f_{123}) 
      +f_3( f_{13} + f_{23} + f_{123}) + f_{123} (f_{12} + f_{13} + f_{23}) \\
    & - f_{123}( f_1 f_{12} + f_1 f_{13} + f_2 f_{12} + f_2 f_{23} + f_3 f_{13} + f_3 f_{23})\Big).
    \label{s4f}
    \end{split}
\end{equation}
At zero temperature it can be expressed in terms of the intersection area of circles as,
\begin{equation}\begin{split}
  s_4 =\frac{1}{(2\pi)^2}\Big(  &  A_1 - \big(A_2({\bf q}_1)+A_2({\bf q}_2)+A_2({\bf q}_3)+A_2({\bf q}_4)+A_2({\bf q}_{12})+A_2({\bf q}_{23})+A_2({\bf q}_{13})\big) \\
    +&2 \big(A_3({\bf q}_1,{\bf q}_2) +A_3({\bf q}_2,{\bf q}_3)+A_3({\bf q}_1,{\bf q}_3)+A_3({\bf q}_1,{\bf q}_4)+A_3({\bf q}_2,{\bf q}_4)+A_3({\bf q}_3,{\bf q}_4)\big)\\
    -&2 \big(A_4({\bf q}_1,{\bf q}_2,{\bf q}_3)+A_4({\bf q}_2,{\bf q}_3,{\bf q}_1)+A_4({\bf q}_3,{\bf q}_1,{\bf q}_2)\big)\Big).
    \label{s4a}
\end{split}\end{equation}
\end{widetext}
Here $A_4({\bf q}_a,{\bf q}_b,{\bf q}_c)$ is the intersection area of four circles centered at $0$, ${\bf q}_a$, ${\bf q}_{ab}$ and ${\bf q}_{abc}$, and we note that
$A_4({\bf q}_a,{\bf q}_b,{\bf q}_c) = A_4({\bf q}_c,{\bf q}_b,{\bf q}_a)$.
Since we have already determined $A_3$, the non-trivial problem is to compute the three terms involving $A_4$.   In two dimensions, $A_4$ can always be expressed in terms of triple intersections $A_3$.

For a given configuration of ${\bf q}_1$, ${\bf q}_2$, ${\bf q}_3$ and ${\bf q}_4 = -{\bf q}_1-{\bf q}_2-{\bf q}_3$, the three terms in Eq. \eqref{s4a} define a sequence of four vectors that when placed end to end
define a 4 sided shape that involves one of each of the patterns depicted in Fig. \ref{fig:4circle}:
(a) a triangle with one point on the inside, (b) a convex quadrilateral, and (c) a figure eight.   
It is always possible to choose $(a,b,c,d)$ to be a permutation of $(1,2,3,4)$, so that the three terms in Eq. \eqref{s4a} correspond to the configurations in Fig. \ref{fig:4circle}.   The appropriate permutation is determined by the sequence of left and right turns, which in turn depend on the signs of $C_{ij}$, defined in Eq. \eqref{cij}.   Note that the 6 numbers that make up $C_{ij}$ are not completely independent.   For example, the fact that $\sum_{p=1}^4 {\bf q}_p = 0$ implies the signs of the non-zero elements in any row or column of $C_{ij}$ are never all the same.    $(a,b,c,d)$ can be chosen such that 
\begin{equation}
{\rm sgn}(C_{ab},C_{bc},C_{cd},C_{da}) = (+1,+1,-1,+1)
\label{c order 1}
\end{equation}
as in Fig. \ref{fig:4circle}(a).    From this, it follows that 
\begin{equation}\begin{split}
{\rm sgn}(C_{ab},C_{bd},C_{dc},C_{ca}) &= (+1,+1,+1,+1)\\
{\rm sgn}(C_{ad},C_{db},C_{bc},C_{ca}) &= (-1,-1,+1,+1)
\end{split}
\label{c order 2}\end{equation}
as in Fig. \ref{fig:4circle}(b,c).    

For each of these cases, the intersection area is determined as follows:

\begin{figure}
    \centering
    \includegraphics[width=.9\linewidth]{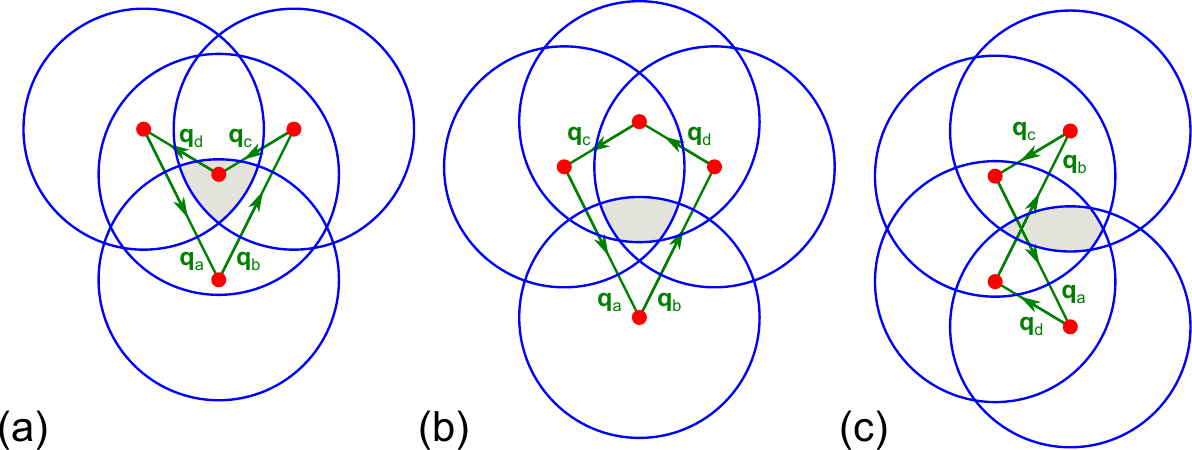}
    \caption{The intersection area of four circles for three configurations.}
    \label{fig:4circle}
\end{figure}

{\it (a)} For the triangle with a point inside, shown in Fig. \ref{fig:4circle}(a), the intersection is determined by the three circles that make the triangle:
\begin{equation}
 A_4({\bf q}_a,{\bf q}_b,{\bf q}_c) = A_3({\bf q}_a,{\bf q}_b) 
 \label{a41}
\end{equation}

{\it (b)} For the convex quadrilateral, shown in Fig. \ref{fig:4circle}(b), the intersection area is determined by adding $A_3$ for the two triangles that share one of the diagonals (which double counts $A_4$), and subtracting $A_2$ associated with that diagonal.    For the configuration in Fig. \ref{fig:4circle}(b), this can be evaluated using either of the two diagonals (i.e.,  ${\bf q}_{ac}$ or ${\bf q}_{ab}$), so there are two equivalent formulas.   However, in some cases, when $A_4=0$, one of these formulas will give an incorrect non-zero result.   The correct one is always the one that involves the triangle with the {\it largest} circle that circumscribes it.    Using the fact that $R({\bf q}_a,{\bf q}_b) = k_{\rm F}/|C_{ab}|$, we therefore have to determine
\begin{equation}\begin{split}
{\cal M}_{ab,dc}&\equiv {\rm min}\{|C_{ab}|,|C_{dc}| \},\\
{\cal M}_{bd,ca}&\equiv {\rm min}\{|C_{bd}|,|C_{ca}| \}.
\end{split}\end{equation}
We then find
\begin{widetext}
\begin{equation}
A_4({\bf q}_c,{\bf q}_a,{\bf q}_b)  = 
   \left\{\begin{array}{ll} A_3({\bf q}_a,{\bf q}_b) + A_3({\bf q}_d,{\bf q}_c) - A_2({\bf q}_{ab})  &  {\rm for} \ 
   {\cal M}_{ab,dc} \le {\cal M}_{bd,ca}  \\
                         A_3({\bf q}_b,{\bf q}_d) + A_3({\bf q}_c,{\bf q}_a) - A_2({\bf q}_{bd})  &  {\rm for} \ 
                      {\cal M}_{ab,dc}   \ge {\cal M}_{bd,ca} .
                         \end{array}\right. 
                         \label{a42}
\end{equation}

{\it (c)} For the figure eight, the analysis is similar to that in {\it (b)} and leads to   
\begin{equation}
 A_4({\bf q}_b,{\bf q}_c,{\bf q}_a)  = 
   \left\{\begin{array}{ll} A_3({\bf q}_a,{\bf q}_d) + A_3({\bf q}_c,{\bf q}_a) - A_2({\bf q}_a)  & {\rm for} \ 
   {\cal M}_{ad,ca} \le {\cal M}_{bc,db} \\
                         A_3({\bf q}_b,{\bf q}_c) + A_3({\bf q}_d,{\bf q}_b) - A_2({\bf q}_b)  &  {\rm for} \ 
                    {\cal M}_{ad,ca}   \ge  {\cal M}_{bc,db} .
                         \end{array}\right.
                         \label{a43}
\end{equation}

For a given ordering of $|C_{ij}|$, it is straightforward to plug the results of Eqs. \eqref{a41}-\eqref{a43} into Eq. \eqref{s4a}.    Using Eq. \eqref{area sum}, we can then express $A_3$ in terms of $s_3$, $A_2$ and $A_1$.   When this is done, we find that the terms involving $A_1$ and $A_2$ all cancel, leading to
\begin{equation}
      s_4({\bf q}_1,{\bf q}_2, {\bf q}_3) = \left\{\begin{array}{ll}
             s_3({\bf q}_a,{\bf q}_c) + s_3({\bf q}_a,{\bf q}_d) - s_3({\bf q}_a,{\bf q}_{b})
       & {\rm for} \   {\cal M}_{ab,dc}\le {\cal M}_{bd,ca} , {\cal M}_{bc,db}\le {\cal M}_{ad,ca}  \\
      s_3({\bf q}_b,{\bf q}_c) + s_3({\bf q}_b,{\bf q}_d) - s_3({\bf q}_b,{\bf q}_a)
       & {\rm for} \  {\cal M}_{ab,dc}\le {\cal M}_{bd,ca} , {\cal M}_{bc,db}\ge {\cal M}_{ad,ca}  \\
      s_3({\bf q}_c,{\bf q}_b) + s_3({\bf q}_c,{\bf q}_d) - s_3({\bf q}_c,{\bf q}_{a})
       & {\rm for} \   {\cal M}_{ab,dc}\ge {\cal M}_{bd,ca} , {\cal M}_{bc,db}\ge {\cal M}_{ad,ca} \\
      s_3({\bf q}_d,{\bf q}_a) + s_3({\bf q}_d,{\bf q}_c) - s_3({\bf q}_d,{\bf q}_{b})
       & {\rm for} \   {\cal M}_{ab,dc}\ge {\cal M}_{bd,ca} , {\cal M}_{bc,db}\le {\cal M}_{ad,ca}  .
    \end{array}\right.
    \label{s4 long}
\end{equation}
\end{widetext}
To apply Eq. \eqref{s4 long} it is necessary to first order ${\bf q}_{a,b,c,d}$ according to Eqs. \eqref{c order 1}, \eqref{c order 2}.   Clearly, the choice of which ${\bf q}$'s appear in the final result that utilizes $s_3$ depends only on the 6 numbers in $C_{ij}$. 
By examining several specific examples, we observed the simpler pattern that is described by the algorithm presented below Eq. \eqref{s4 formula} (which does not require ordering the ${\bf q}$'s).  
We have proven that this algorithm is equivalent to Eq. \eqref{s4 long} by exhaustively checking all of the possible orderings of $|C_{ij}|$ and utilizing certain constraints obeyed by $C_{ij}$.   The proof is long and tedious, and will not be reproduced here.   An alternative demonstration of the validity of the algorithm is to consider a large ensemble of randomly chosen configurations of ${\bf q}_{1,2,3}$, and check that the algorithm agrees with the expression obtained from Eq. \eqref{s4 long}, as well as with the numerical integration of Eq. \eqref{s4f}.   We have checked that it works in every case.

\cleardoublepage

\onecolumngrid
\subsection*{Supplementary Figures}

\begin{figure*}[h!]
    \centering
    \includegraphics[width = 0.5\textwidth]{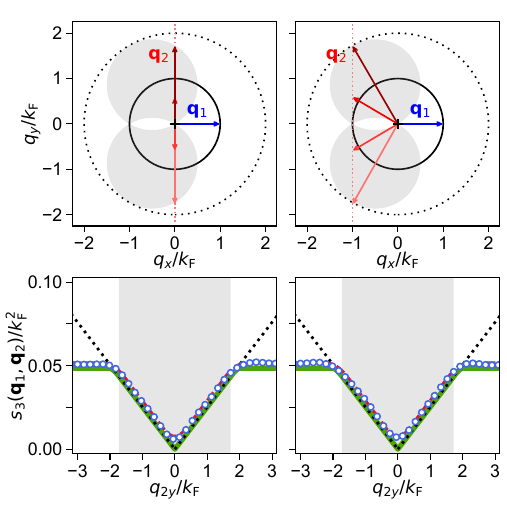}
    \caption{\textbf{$s_3$ from two triangulations of the Fermi sea spanning the same set of vertices.} $\textbf{q}_1$ is fixed. We fix the projection of $\textbf{q}_2$ along $\textbf{q}_1$ and vary its orthogonal component $q_{2y}$. (Left) $|\textbf{q}_1|/k_{\rm F}=1$, $q_{2x}/k_{\rm F}=0$; (Right) $|\textbf{q}_1|/k_{\rm F}=1$, $q_{2x}/k_{\rm F}=-1$. These two configurations lead to triangulations of the Fermi sea spanning the same set of vertices and mirroring one another. This leads to the same $s_3$ within experimental errors (Bottom row) and further highlights the volume scaling of $s_3$. Blue markers: Experimental data with ${\cal I} = -0.129(3)$; Green curve: Analytical result; red curve: numerical results. 
    }
    \label{fig:figS1}
\end{figure*}


\begin{figure*}
    \centering
    \includegraphics[width = \textwidth]{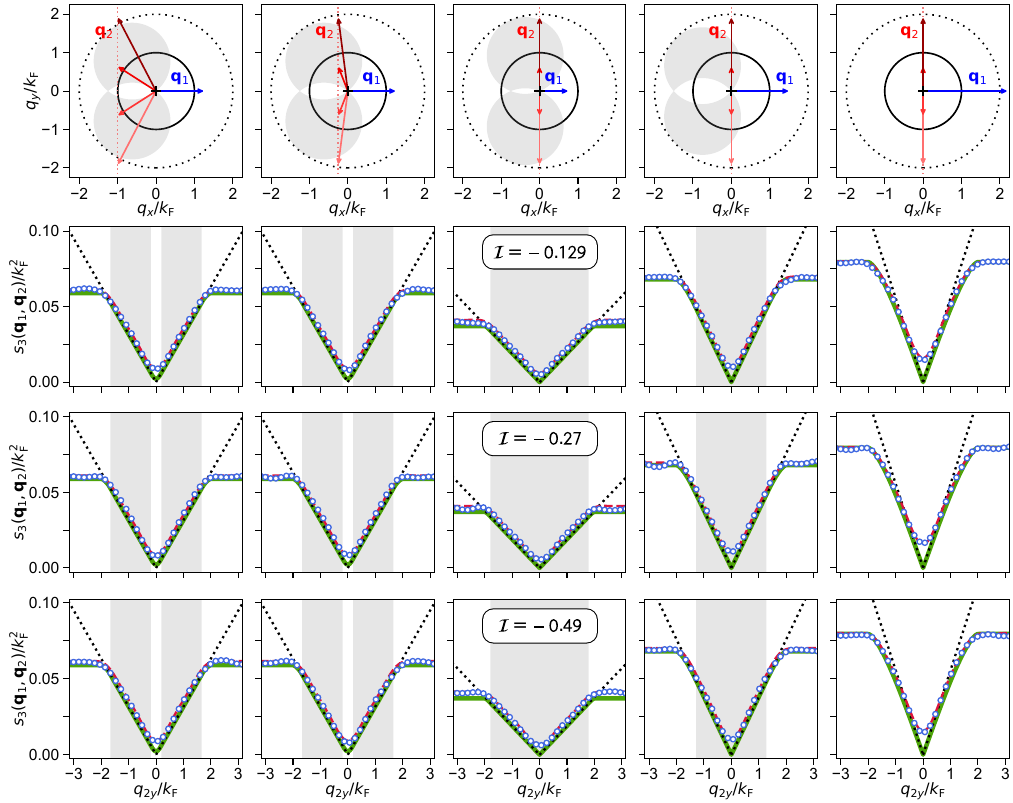}
    \caption{\textbf{$s_3$ for different interaction strengths}. Top row: Selected configurations $\{\textbf{q}_a\}$. $\textbf{q}_1$ is fixed while $\textbf{q}_2$ is taken along the dotted red line. From left to right: $|\textbf{q}_1|/k_{\rm F}=1.25$, $q_{2x}/k_{\rm F}=-1$; $|\textbf{q}_1|/k_{\rm F}=1.25$, $q_{2x}/k_{\rm F}=-0.25$; $|\textbf{q}_1|/k_{\rm F}=0.75$, $q_{2x}/k_{\rm F}=0$; $|\textbf{q}_1|/k_{\rm F}=1.5$, $q_{2x}/k_{\rm F}=0$; $|\textbf{q}_1|/k_{\rm F}=2.20$, $q_{2x}/k_{\rm F}=-0$.
    From top to bottom: ${\cal I}= -0.129(3)$,  $-0.27(1)$, and ${\cal I}=-0.49(2)$, corresponding to $T/T_{\rm F}=0.15(1), 0.15(1), 0.14(2)$ and $p_1 = 4.7(4)\%,\ 4.3(3)\%,\ 5.6(4)\%$ respectively. In this interaction regime, $s_3$ is seen to remain unmodified by interactions. The rightmost pannel showcases the breakdown of the topological formula for large values of the triangulating vectors.}
    \label{fig:figS5}
\end{figure*}

\begin{figure*}
    \centering
    \includegraphics[width = \textwidth]{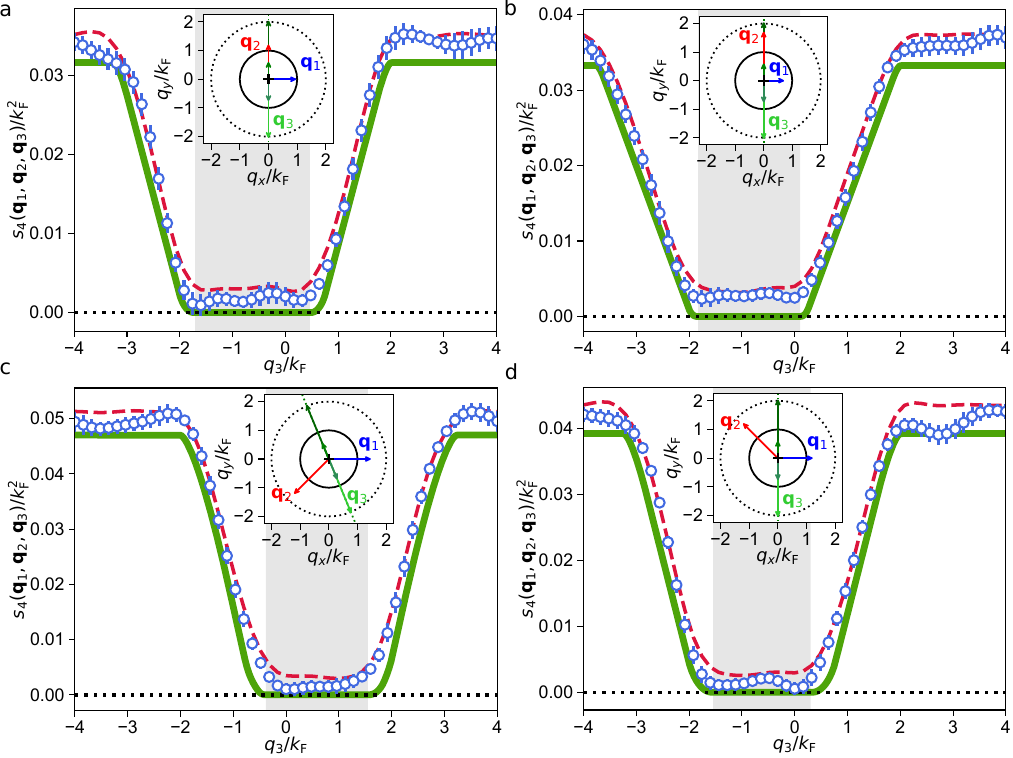}
    \caption{\textbf{Four-point connected density correlations.} $s_4$ for various configurations $\{\textbf{q}_p \}$ (Insets). $\textbf{q}_1$ and $\textbf{q}_2$ are kept constant while $\textbf{q}_3$ is varied along the dotted green line. Blue markers: Experimental data with ${\cal I} =-0.129(3)$; Green curve: Analytical result; red curve: numerical results. (a) $|\textbf{q}_1|/k_{\rm F}=1$, $|\textbf{q}_2|/k_{\rm F}=1.25$, $\theta_{12}=\pi/2$, $\theta_{13}=\theta_{12}\ \rm{mod}\ \pi$; (b) $|\textbf{q}_1|/k_{\rm F}=0.75$, $|\textbf{q}_2|/k_{\rm F}=1.75$, $\theta_{12}=\pi/2$, $\theta_{13}=\theta_{12}\ \rm{mod}\ \pi$; (c) $|\textbf{q}_1|/k_{\rm F}=1.5$, $|\textbf{q}_2|/k_{\rm F}=1.75$, $\theta_{12}=5\pi/4$, $\theta_{13}=\theta_{12}/2\ \rm{mod}\ \pi$; (d) $|\textbf{q}_1|/k_{\rm F}=1.25$, $|\textbf{q}_2|/k_{\rm F}=1.75$, $\theta_{12}=3\pi/4$, $\theta_{13}=\pi/2\ \rm{mod}\ \pi$. $s_4$ is seen to remain constant and close to zero over a large region in momentum space, which coincides with the region of validity for $s_4=0$ (grey area).}
    \label{fig:figS3}
\end{figure*}


\begin{thebibliography}{37}%
\makeatletter
\providecommand \@ifxundefined [1]{%
 \@ifx{#1\undefined}
}%
\providecommand \@ifnum [1]{%
 \ifnum #1\expandafter \@firstoftwo
 \else \expandafter \@secondoftwo
 \fi
}%
\providecommand \@ifx [1]{%
 \ifx #1\expandafter \@firstoftwo
 \else \expandafter \@secondoftwo
 \fi
}%
\providecommand \natexlab [1]{#1}%
\providecommand \enquote  [1]{``#1''}%
\providecommand \bibnamefont  [1]{#1}%
\providecommand \bibfnamefont [1]{#1}%
\providecommand \citenamefont [1]{#1}%
\providecommand \href@noop [0]{\@secondoftwo}%
\providecommand \href [0]{\begingroup \@sanitize@url \@href}%
\providecommand \@href[1]{\@@startlink{#1}\@@href}%
\providecommand \@@href[1]{\endgroup#1\@@endlink}%
\providecommand \@sanitize@url [0]{\catcode `\\12\catcode `\$12\catcode
  `\&12\catcode `\#12\catcode `\^12\catcode `\_12\catcode `\%12\relax}%
\providecommand \@@startlink[1]{}%
\providecommand \@@endlink[0]{}%
\providecommand \url  [0]{\begingroup\@sanitize@url \@url }%
\providecommand \@url [1]{\endgroup\@href {#1}{\urlprefix }}%
\providecommand \urlprefix  [0]{URL }%
\providecommand \Eprint [0]{\href }%
\providecommand \doibase [0]{https://doi.org/}%
\providecommand \selectlanguage [0]{\@gobble}%
\providecommand \bibinfo  [0]{\@secondoftwo}%
\providecommand \bibfield  [0]{\@secondoftwo}%
\providecommand \translation [1]{[#1]}%
\providecommand \BibitemOpen [0]{}%
\providecommand \bibitemStop [0]{}%
\providecommand \bibitemNoStop [0]{.\EOS\space}%
\providecommand \EOS [0]{\spacefactor3000\relax}%
\providecommand \BibitemShut  [1]{\csname bibitem#1\endcsname}%
\let\auto@bib@innerbib\@empty
\bibitem [{\citenamefont {Kane}(2022)}]{kane2022}%
  \BibitemOpen
  \bibfield  {author} {\bibinfo {author} {\bibfnamefont {C.~L.}\ \bibnamefont
  {Kane}},\ }\href {https://doi.org/10.1103/PhysRevLett.128.076801} {\bibfield
  {journal} {\bibinfo  {journal} {Physical Review Letters}\ }\textbf {\bibinfo
  {volume} {128}},\ \bibinfo {pages} {076801} (\bibinfo {year}
  {2022})}\BibitemShut {NoStop}%
\bibitem [{\citenamefont {Van~Wees}\ \emph {et~al.}(1988)\citenamefont
  {Van~Wees}, \citenamefont {Van~Houten}, \citenamefont {Beenakker},
  \citenamefont {Williamson}, \citenamefont {Kouwenhoven}, \citenamefont {Van
  Der~Marel},\ and\ \citenamefont {Foxon}}]{vanwees1988}%
  \BibitemOpen
  \bibfield  {author} {\bibinfo {author} {\bibfnamefont {B.~J.}\ \bibnamefont
  {Van~Wees}}, \bibinfo {author} {\bibfnamefont {H.}~\bibnamefont
  {Van~Houten}}, \bibinfo {author} {\bibfnamefont {C.~W.~J.}\ \bibnamefont
  {Beenakker}}, \bibinfo {author} {\bibfnamefont {J.~G.}\ \bibnamefont
  {Williamson}}, \bibinfo {author} {\bibfnamefont {L.~P.}\ \bibnamefont
  {Kouwenhoven}}, \bibinfo {author} {\bibfnamefont {D.}~\bibnamefont {Van
  Der~Marel}},\ and\ \bibinfo {author} {\bibfnamefont {C.~T.}\ \bibnamefont
  {Foxon}},\ }\href {https://doi.org/10.1103/PhysRevLett.60.848} {\bibfield
  {journal} {\bibinfo  {journal} {Physical Review Letters}\ }\textbf {\bibinfo
  {volume} {60}},\ \bibinfo {pages} {848} (\bibinfo {year} {1988})}\BibitemShut
  {NoStop}%
\bibitem [{\citenamefont {Honda}\ \emph {et~al.}(1995)\citenamefont {Honda},
  \citenamefont {Tarucha}, \citenamefont {Saku},\ and\ \citenamefont
  {Yasuhiro~Tokura}}]{honda1995}%
  \BibitemOpen
  \bibfield  {author} {\bibinfo {author} {\bibfnamefont {T.}~\bibnamefont
  {Honda}}, \bibinfo {author} {\bibfnamefont {S.}~\bibnamefont {Tarucha}},
  \bibinfo {author} {\bibfnamefont {T.}~\bibnamefont {Saku}},\ and\ \bibinfo
  {author} {\bibfnamefont {Y.~T.}\ \bibnamefont {Yasuhiro~Tokura}},\ }\href
  {https://doi.org/10.1143/JJAP.34.L72} {\bibfield  {journal} {\bibinfo
  {journal} {Japanese Journal of Applied Physics}\ }\textbf {\bibinfo {volume}
  {34}},\ \bibinfo {pages} {L72} (\bibinfo {year} {1995})}\BibitemShut
  {NoStop}%
\bibitem [{\citenamefont {Van~Weperen}\ \emph {et~al.}(2013)\citenamefont
  {Van~Weperen}, \citenamefont {Plissard}, \citenamefont {Bakkers},
  \citenamefont {Frolov},\ and\ \citenamefont {Kouwenhoven}}]{vanweperen2013}%
  \BibitemOpen
  \bibfield  {author} {\bibinfo {author} {\bibfnamefont {I.}~\bibnamefont
  {Van~Weperen}}, \bibinfo {author} {\bibfnamefont {S.~R.}\ \bibnamefont
  {Plissard}}, \bibinfo {author} {\bibfnamefont {E.~P. A.~M.}\ \bibnamefont
  {Bakkers}}, \bibinfo {author} {\bibfnamefont {S.~M.}\ \bibnamefont
  {Frolov}},\ and\ \bibinfo {author} {\bibfnamefont {L.~P.}\ \bibnamefont
  {Kouwenhoven}},\ }\href {https://doi.org/10.1021/nl3035256} {\bibfield
  {journal} {\bibinfo  {journal} {Nano Letters}\ }\textbf {\bibinfo {volume}
  {13}},\ \bibinfo {pages} {387} (\bibinfo {year} {2013})}\BibitemShut
  {NoStop}%
\bibitem [{\citenamefont {Frank}\ \emph {et~al.}(1998)\citenamefont {Frank},
  \citenamefont {Poncharal}, \citenamefont {Wang},\ and\ \citenamefont
  {Heer}}]{frank1998}%
  \BibitemOpen
  \bibfield  {author} {\bibinfo {author} {\bibfnamefont {S.}~\bibnamefont
  {Frank}}, \bibinfo {author} {\bibfnamefont {P.}~\bibnamefont {Poncharal}},
  \bibinfo {author} {\bibfnamefont {Z.~L.}\ \bibnamefont {Wang}},\ and\
  \bibinfo {author} {\bibfnamefont {W.~A.~D.}\ \bibnamefont {Heer}},\ }\href
  {https://doi.org/10.1126/science.280.5370.1744} {\bibfield  {journal}
  {\bibinfo  {journal} {Science}\ }\textbf {\bibinfo {volume} {280}},\ \bibinfo
  {pages} {1744} (\bibinfo {year} {1998})}\BibitemShut {NoStop}%
\bibitem [{\citenamefont {Krinner}\ \emph {et~al.}(2015)\citenamefont
  {Krinner}, \citenamefont {Stadler}, \citenamefont {Husmann}, \citenamefont
  {Brantut},\ and\ \citenamefont {Esslinger}}]{krinner2015}%
  \BibitemOpen
  \bibfield  {author} {\bibinfo {author} {\bibfnamefont {S.}~\bibnamefont
  {Krinner}}, \bibinfo {author} {\bibfnamefont {D.}~\bibnamefont {Stadler}},
  \bibinfo {author} {\bibfnamefont {D.}~\bibnamefont {Husmann}}, \bibinfo
  {author} {\bibfnamefont {J.-P.}\ \bibnamefont {Brantut}},\ and\ \bibinfo
  {author} {\bibfnamefont {T.}~\bibnamefont {Esslinger}},\ }\href
  {https://doi.org/10.1038/nature14049} {\bibfield  {journal} {\bibinfo
  {journal} {Nature}\ }\textbf {\bibinfo {volume} {517}},\ \bibinfo {pages}
  {64} (\bibinfo {year} {2015})}\BibitemShut {NoStop}%
\bibitem [{\citenamefont {Klitzing}\ \emph {et~al.}(1980)\citenamefont
  {Klitzing}, \citenamefont {Dorda},\ and\ \citenamefont
  {Pepper}}]{klitzing1980}%
  \BibitemOpen
  \bibfield  {author} {\bibinfo {author} {\bibfnamefont {K.~V.}\ \bibnamefont
  {Klitzing}}, \bibinfo {author} {\bibfnamefont {G.}~\bibnamefont {Dorda}},\
  and\ \bibinfo {author} {\bibfnamefont {M.}~\bibnamefont {Pepper}},\ }\href
  {https://doi.org/10.1103/PhysRevLett.45.494} {\bibfield  {journal} {\bibinfo
  {journal} {Physical Review Letters}\ }\textbf {\bibinfo {volume} {45}},\
  \bibinfo {pages} {494} (\bibinfo {year} {1980})}\BibitemShut {NoStop}%
\bibitem [{\citenamefont {Thouless}\ \emph {et~al.}(1982)\citenamefont
  {Thouless}, \citenamefont {Kohmoto}, \citenamefont {Nightingale},\ and\
  \citenamefont {Den~Nijs}}]{thouless1982}%
  \BibitemOpen
  \bibfield  {author} {\bibinfo {author} {\bibfnamefont {D.~J.}\ \bibnamefont
  {Thouless}}, \bibinfo {author} {\bibfnamefont {M.}~\bibnamefont {Kohmoto}},
  \bibinfo {author} {\bibfnamefont {M.~P.}\ \bibnamefont {Nightingale}},\ and\
  \bibinfo {author} {\bibfnamefont {M.}~\bibnamefont {Den~Nijs}},\ }\href
  {https://doi.org/10.1103/PhysRevLett.49.405} {\bibfield  {journal} {\bibinfo
  {journal} {Physical Review Letters}\ }\textbf {\bibinfo {volume} {49}},\
  \bibinfo {pages} {405} (\bibinfo {year} {1982})}\BibitemShut {NoStop}%
\bibitem [{\citenamefont {Sanner}\ \emph {et~al.}(2021)\citenamefont {Sanner},
  \citenamefont {Sonderhouse}, \citenamefont {Hutson}, \citenamefont {Yan},
  \citenamefont {Milner},\ and\ \citenamefont {Ye}}]{sanner2021}%
  \BibitemOpen
  \bibfield  {author} {\bibinfo {author} {\bibfnamefont {C.}~\bibnamefont
  {Sanner}}, \bibinfo {author} {\bibfnamefont {L.}~\bibnamefont {Sonderhouse}},
  \bibinfo {author} {\bibfnamefont {R.~B.}\ \bibnamefont {Hutson}}, \bibinfo
  {author} {\bibfnamefont {L.}~\bibnamefont {Yan}}, \bibinfo {author}
  {\bibfnamefont {W.~R.}\ \bibnamefont {Milner}},\ and\ \bibinfo {author}
  {\bibfnamefont {J.}~\bibnamefont {Ye}},\ }\href
  {https://doi.org/10.1126/science.abh3483} {\bibfield  {journal} {\bibinfo
  {journal} {Science}\ }\textbf {\bibinfo {volume} {374}},\ \bibinfo {pages}
  {979} (\bibinfo {year} {2021})}\BibitemShut {NoStop}%
\bibitem [{\citenamefont {Deb}\ and\ \citenamefont
  {Kj{\ae}rgaard}(2021)}]{deb2021}%
  \BibitemOpen
  \bibfield  {author} {\bibinfo {author} {\bibfnamefont {A.~B.}\ \bibnamefont
  {Deb}}\ and\ \bibinfo {author} {\bibfnamefont {N.}~\bibnamefont
  {Kj{\ae}rgaard}},\ }\href {https://doi.org/10.1126/science.abh3470}
  {\bibfield  {journal} {\bibinfo  {journal} {Science}\ }\textbf {\bibinfo
  {volume} {374}},\ \bibinfo {pages} {972} (\bibinfo {year}
  {2021})}\BibitemShut {NoStop}%
\bibitem [{\citenamefont {Margalit}\ \emph {et~al.}(2021)\citenamefont
  {Margalit}, \citenamefont {Lu}, \citenamefont {Top},\ and\ \citenamefont
  {Ketterle}}]{margalit2021}%
  \BibitemOpen
  \bibfield  {author} {\bibinfo {author} {\bibfnamefont {Y.}~\bibnamefont
  {Margalit}}, \bibinfo {author} {\bibfnamefont {Y.-K.}\ \bibnamefont {Lu}},
  \bibinfo {author} {\bibfnamefont {F.~{\c C}.}\ \bibnamefont {Top}},\ and\
  \bibinfo {author} {\bibfnamefont {W.}~\bibnamefont {Ketterle}},\ }\href
  {https://doi.org/10.1126/science.abi6153} {\bibfield  {journal} {\bibinfo
  {journal} {Science}\ }\textbf {\bibinfo {volume} {374}},\ \bibinfo {pages}
  {976} (\bibinfo {year} {2021})}\BibitemShut {NoStop}%
\bibitem [{\citenamefont {Jannin}\ \emph {et~al.}(2022)\citenamefont {Jannin},
  \citenamefont {Van Der~Werf}, \citenamefont {Steinebach}, \citenamefont
  {Bethlem},\ and\ \citenamefont {Eikema}}]{jannin2022}%
  \BibitemOpen
  \bibfield  {author} {\bibinfo {author} {\bibfnamefont {R.}~\bibnamefont
  {Jannin}}, \bibinfo {author} {\bibfnamefont {Y.}~\bibnamefont {Van
  Der~Werf}}, \bibinfo {author} {\bibfnamefont {K.}~\bibnamefont {Steinebach}},
  \bibinfo {author} {\bibfnamefont {H.~L.}\ \bibnamefont {Bethlem}},\ and\
  \bibinfo {author} {\bibfnamefont {K.~S.~E.}\ \bibnamefont {Eikema}},\ }\href
  {https://doi.org/10.1038/s41467-022-34135-6} {\bibfield  {journal} {\bibinfo
  {journal} {Nature Communications}\ }\textbf {\bibinfo {volume} {13}},\
  \bibinfo {pages} {6479} (\bibinfo {year} {2022})}\BibitemShut {NoStop}%
\bibitem [{\citenamefont {Zwierlein}(2013)}]{zwierlein2013}%
  \BibitemOpen
  \bibfield  {author} {\bibinfo {author} {\bibfnamefont {M.~W.}\ \bibnamefont
  {Zwierlein}},\ }in\ \href@noop {} {\emph {\bibinfo {booktitle} {Novel
  {{Superfluids}}}}},\ Vol.~\bibinfo {volume} {II},\ \bibinfo {editor} {edited
  by\ \bibinfo {editor} {\bibfnamefont {K.}~\bibnamefont {Bennemann}}\ and\
  \bibinfo {editor} {\bibfnamefont {J.}~\bibnamefont {Ketterson}}}\ (\bibinfo
  {publisher} {Oxford University Press},\ \bibinfo {address} {Oxford, UK},\
  \bibinfo {year} {2013})\ pp.\ \bibinfo {pages} {269--422}\BibitemShut
  {NoStop}%
\bibitem [{\citenamefont {Holten}\ \emph {et~al.}(2022)\citenamefont {Holten},
  \citenamefont {Bayha}, \citenamefont {Subramanian}, \citenamefont
  {Brandstetter}, \citenamefont {Heintze}, \citenamefont {Lunt}, \citenamefont
  {Preiss},\ and\ \citenamefont {Jochim}}]{holten_2022}%
  \BibitemOpen
  \bibfield  {author} {\bibinfo {author} {\bibfnamefont {M.}~\bibnamefont
  {Holten}}, \bibinfo {author} {\bibfnamefont {L.}~\bibnamefont {Bayha}},
  \bibinfo {author} {\bibfnamefont {K.}~\bibnamefont {Subramanian}}, \bibinfo
  {author} {\bibfnamefont {S.}~\bibnamefont {Brandstetter}}, \bibinfo {author}
  {\bibfnamefont {C.}~\bibnamefont {Heintze}}, \bibinfo {author} {\bibfnamefont
  {P.}~\bibnamefont {Lunt}}, \bibinfo {author} {\bibfnamefont {P.~M.}\
  \bibnamefont {Preiss}},\ and\ \bibinfo {author} {\bibfnamefont
  {S.}~\bibnamefont {Jochim}},\ }\href
  {https://doi.org/10.1038/s41586-022-04678-1} {\bibfield  {journal} {\bibinfo
  {journal} {Nature}\ }\textbf {\bibinfo {volume} {606}},\ \bibinfo {pages}
  {287} (\bibinfo {year} {2022})}\BibitemShut {NoStop}%
\bibitem [{\citenamefont {Yang}\ and\ \citenamefont {Zhai}(2022)}]{yang2022}%
  \BibitemOpen
  \bibfield  {author} {\bibinfo {author} {\bibfnamefont {F.}~\bibnamefont
  {Yang}}\ and\ \bibinfo {author} {\bibfnamefont {H.}~\bibnamefont {Zhai}},\
  }\href {https://doi.org/10.22331/q-2022-11-10-857} {\bibfield  {journal}
  {\bibinfo  {journal} {Quantum}\ }\textbf {\bibinfo {volume} {6}},\ \bibinfo
  {pages} {857} (\bibinfo {year} {2022})}\BibitemShut {NoStop}%
\bibitem [{\citenamefont {Zhang}(2023)}]{zhang2023}%
  \BibitemOpen
  \bibfield  {author} {\bibinfo {author} {\bibfnamefont {P.}~\bibnamefont
  {Zhang}},\ }\href {https://doi.org/10.1103/PhysRevA.107.L031305} {\bibfield
  {journal} {\bibinfo  {journal} {Physical Review A}\ }\textbf {\bibinfo
  {volume} {107}},\ \bibinfo {pages} {L031305} (\bibinfo {year}
  {2023})}\BibitemShut {NoStop}%
\bibitem [{\citenamefont {Tam}\ and\ \citenamefont {Kane}(2023)}]{tam2023}%
  \BibitemOpen
  \bibfield  {author} {\bibinfo {author} {\bibfnamefont {P.~M.}\ \bibnamefont
  {Tam}}\ and\ \bibinfo {author} {\bibfnamefont {C.~L.}\ \bibnamefont {Kane}},\
  }\href {https://doi.org/10.1103/PhysRevLett.130.096301} {\bibfield  {journal}
  {\bibinfo  {journal} {Physical Review Letters}\ }\textbf {\bibinfo {volume}
  {130}},\ \bibinfo {pages} {096301} (\bibinfo {year} {2023})}\BibitemShut
  {NoStop}%
\bibitem [{\citenamefont {Tam}\ \emph {et~al.}(2023)\citenamefont {Tam},
  \citenamefont {De~Beule},\ and\ \citenamefont {Kane}}]{tam2023a}%
  \BibitemOpen
  \bibfield  {author} {\bibinfo {author} {\bibfnamefont {P.~M.}\ \bibnamefont
  {Tam}}, \bibinfo {author} {\bibfnamefont {C.}~\bibnamefont {De~Beule}},\ and\
  \bibinfo {author} {\bibfnamefont {C.~L.}\ \bibnamefont {Kane}},\ }\href
  {https://doi.org/10.1103/PhysRevB.107.245422} {\bibfield  {journal} {\bibinfo
   {journal} {Physical Review B}\ }\textbf {\bibinfo {volume} {107}},\ \bibinfo
  {pages} {245422} (\bibinfo {year} {2023})}\BibitemShut {NoStop}%
\bibitem [{\citenamefont {Tam}\ \emph {et~al.}(2022)\citenamefont {Tam},
  \citenamefont {Claassen},\ and\ \citenamefont {Kane}}]{tam2022}%
  \BibitemOpen
  \bibfield  {author} {\bibinfo {author} {\bibfnamefont {P.~M.}\ \bibnamefont
  {Tam}}, \bibinfo {author} {\bibfnamefont {M.}~\bibnamefont {Claassen}},\ and\
  \bibinfo {author} {\bibfnamefont {C.~L.}\ \bibnamefont {Kane}},\ }\href
  {https://doi.org/10.1103/PhysRevX.12.031022} {\bibfield  {journal} {\bibinfo
  {journal} {Physical Review X}\ }\textbf {\bibinfo {volume} {12}},\ \bibinfo
  {pages} {031022} (\bibinfo {year} {2022})}\BibitemShut {NoStop}%
\bibitem [{\citenamefont {Tam}\ and\ \citenamefont {Kane}(2024)}]{tam2024}%
  \BibitemOpen
  \bibfield  {author} {\bibinfo {author} {\bibfnamefont {P.~M.}\ \bibnamefont
  {Tam}}\ and\ \bibinfo {author} {\bibfnamefont {C.~L.}\ \bibnamefont {Kane}},\
  }\href {https://doi.org/10.1103/PhysRevB.109.035413} {\bibfield  {journal}
  {\bibinfo  {journal} {Physical Review B}\ }\textbf {\bibinfo {volume}
  {109}},\ \bibinfo {pages} {035413} (\bibinfo {year} {2024})}\BibitemShut
  {NoStop}%
\bibitem [{\citenamefont {Euler}(1758)}]{euler1758}%
  \BibitemOpen
  \bibfield  {author} {\bibinfo {author} {\bibfnamefont {L.}~\bibnamefont
  {Euler}},\ }\href@noop {} {\bibfield  {journal} {\bibinfo  {journal} {Novi
  Commentarii academiae scientiarum Petropolitanae}\ }\textbf {\bibinfo
  {volume} {4}},\ \bibinfo {pages} {109} (\bibinfo {year} {1758})}\BibitemShut
  {NoStop}%
\bibitem [{\citenamefont {Castin}(2007)}]{castin2007}%
  \BibitemOpen
  \bibfield  {author} {\bibinfo {author} {\bibfnamefont {Y.}~\bibnamefont
  {Castin}},\ }in\ \href {https://doi.org/10.3254/978-1-58603-846-5-289} {\emph
  {\bibinfo {booktitle} {Lecture Notes of the 2006 {{Varenna Enrico Fermi
  School}} on {{Fermi}} Gases}}},\ \bibinfo {editor} {edited by\ \bibinfo
  {editor} {\bibfnamefont {M.}~\bibnamefont {Inguscio}}, \bibinfo {editor}
  {\bibfnamefont {W.}~\bibnamefont {Ketterle}},\ and\ \bibinfo {editor}
  {\bibfnamefont {C.}~\bibnamefont {Salomon}}}\ (\bibinfo  {publisher} {IOS
  Press},\ \bibinfo {year} {2007})\ pp.\ \bibinfo {pages}
  {289--349}\BibitemShut {NoStop}%
\bibitem [{\citenamefont {Levinsen}\ and\ \citenamefont
  {Parish}(2015)}]{levinsen2015}%
  \BibitemOpen
  \bibfield  {author} {\bibinfo {author} {\bibfnamefont {J.}~\bibnamefont
  {Levinsen}}\ and\ \bibinfo {author} {\bibfnamefont {M.~M.}\ \bibnamefont
  {Parish}},\ }in\ \href {https://doi.org/10.1142/9789814667746_0001} {\emph
  {\bibinfo {booktitle} {Annual {{Review}} of {{Cold Atoms}} and
  {{Molecules}}}}},\ \bibinfo {series} {Annual {{Review}} of {{Cold Atoms}} and
  {{Molecules}}}, Vol.\ \bibinfo {volume} {Volume 3}\ (\bibinfo  {publisher}
  {WORLD SCIENTIFIC},\ \bibinfo {year} {2015})\ pp.\ \bibinfo {pages}
  {1--75}\BibitemShut {NoStop}%
\bibitem [{\citenamefont {Daix}\ \emph {et~al.}(2025)\citenamefont {Daix},
  \citenamefont {Dixmerias}, \citenamefont {He}, \citenamefont {Verstraten},
  \citenamefont {de~Jongh}, \citenamefont {Peaudecerf}, \citenamefont {Zhang},\
  and\ \citenamefont {Yefsah}}]{daix2025}%
  \BibitemOpen
  \bibfield  {author} {\bibinfo {author} {\bibfnamefont {C.}~\bibnamefont
  {Daix}}, \bibinfo {author} {\bibfnamefont {M.}~\bibnamefont {Dixmerias}},
  \bibinfo {author} {\bibfnamefont {Y.-Y.}\ \bibnamefont {He}}, \bibinfo
  {author} {\bibfnamefont {J.}~\bibnamefont {Verstraten}}, \bibinfo {author}
  {\bibfnamefont {T.}~\bibnamefont {de~Jongh}}, \bibinfo {author}
  {\bibfnamefont {B.}~\bibnamefont {Peaudecerf}}, \bibinfo {author}
  {\bibfnamefont {S.}~\bibnamefont {Zhang}},\ and\ \bibinfo {author}
  {\bibfnamefont {T.}~\bibnamefont {Yefsah}},\ }\href
  {https://doi.org/10.48550/arXiv.2504.01885} {\bibinfo {title} {Observing
  {{Spatial Charge}} and {{Spin Correlations}} in a {{Strongly-Interacting
  Fermi Gas}}}} (\bibinfo {year} {2025}),\ \Eprint
  {https://arxiv.org/abs/2504.01885} {arXiv:2504.01885 [cond-mat]} \BibitemShut
  {NoStop}%
\bibitem [{sup()}]{supmat}%
  \BibitemOpen
  \href@noop {} {}\bibinfo {note} {See Supplementary Materials}\BibitemShut
  {NoStop}%
\bibitem [{\citenamefont {Verstraten}\ \emph {et~al.}(2025)\citenamefont
  {Verstraten}, \citenamefont {Dai}, \citenamefont {Dixmerias}, \citenamefont
  {Peaudecerf}, \citenamefont {{de Jongh}},\ and\ \citenamefont
  {Yefsah}}]{verstraten2025}%
  \BibitemOpen
  \bibfield  {author} {\bibinfo {author} {\bibfnamefont {J.}~\bibnamefont
  {Verstraten}}, \bibinfo {author} {\bibfnamefont {K.}~\bibnamefont {Dai}},
  \bibinfo {author} {\bibfnamefont {M.}~\bibnamefont {Dixmerias}}, \bibinfo
  {author} {\bibfnamefont {B.}~\bibnamefont {Peaudecerf}}, \bibinfo {author}
  {\bibfnamefont {T.}~\bibnamefont {{de Jongh}}},\ and\ \bibinfo {author}
  {\bibfnamefont {T.}~\bibnamefont {Yefsah}},\ }\href
  {https://doi.org/10.1103/PhysRevLett.134.083403} {\bibfield  {journal}
  {\bibinfo  {journal} {Physical Review Letters}\ }\textbf {\bibinfo {volume}
  {134}},\ \bibinfo {pages} {083403} (\bibinfo {year} {2025})}\BibitemShut
  {NoStop}%
\bibitem [{\citenamefont {{de Jongh}}\ \emph {et~al.}(2025)\citenamefont {{de
  Jongh}}, \citenamefont {Verstraten}, \citenamefont {Dixmerias}, \citenamefont
  {Daix}, \citenamefont {Peaudecerf},\ and\ \citenamefont
  {Yefsah}}]{dejongh2025}%
  \BibitemOpen
  \bibfield  {author} {\bibinfo {author} {\bibfnamefont {T.}~\bibnamefont {{de
  Jongh}}}, \bibinfo {author} {\bibfnamefont {J.}~\bibnamefont {Verstraten}},
  \bibinfo {author} {\bibfnamefont {M.}~\bibnamefont {Dixmerias}}, \bibinfo
  {author} {\bibfnamefont {C.}~\bibnamefont {Daix}}, \bibinfo {author}
  {\bibfnamefont {B.}~\bibnamefont {Peaudecerf}},\ and\ \bibinfo {author}
  {\bibfnamefont {T.}~\bibnamefont {Yefsah}},\ }\href
  {https://doi.org/10.1103/PhysRevLett.134.183403} {\bibfield  {journal}
  {\bibinfo  {journal} {Physical Review Letters}\ }\textbf {\bibinfo {volume}
  {134}},\ \bibinfo {pages} {183403} (\bibinfo {year} {2025})}\BibitemShut
  {NoStop}%
\bibitem [{\citenamefont {Yao}\ \emph {et~al.}(2025)\citenamefont {Yao},
  \citenamefont {Chi}, \citenamefont {Wang}, \citenamefont {Fletcher},\ and\
  \citenamefont {Zwierlein}}]{yao2025}%
  \BibitemOpen
  \bibfield  {author} {\bibinfo {author} {\bibfnamefont {R.}~\bibnamefont
  {Yao}}, \bibinfo {author} {\bibfnamefont {S.}~\bibnamefont {Chi}}, \bibinfo
  {author} {\bibfnamefont {M.}~\bibnamefont {Wang}}, \bibinfo {author}
  {\bibfnamefont {R.~J.}\ \bibnamefont {Fletcher}},\ and\ \bibinfo {author}
  {\bibfnamefont {M.}~\bibnamefont {Zwierlein}},\ }\href
  {https://doi.org/10.1103/PhysRevLett.134.183402} {\bibfield  {journal}
  {\bibinfo  {journal} {Physical Review Letters}\ }\textbf {\bibinfo {volume}
  {134}},\ \bibinfo {pages} {183402} (\bibinfo {year} {2025})}\BibitemShut
  {NoStop}%
\bibitem [{\citenamefont {Xiang}\ \emph {et~al.}(2025)\citenamefont {Xiang},
  \citenamefont {{Cruz-Col{\'o}n}}, \citenamefont {Chua}, \citenamefont
  {Milner}, \citenamefont {De~Hond}, \citenamefont {Fricke},\ and\
  \citenamefont {Ketterle}}]{xiang2025}%
  \BibitemOpen
  \bibfield  {author} {\bibinfo {author} {\bibfnamefont {J.}~\bibnamefont
  {Xiang}}, \bibinfo {author} {\bibfnamefont {E.}~\bibnamefont
  {{Cruz-Col{\'o}n}}}, \bibinfo {author} {\bibfnamefont {C.~C.}\ \bibnamefont
  {Chua}}, \bibinfo {author} {\bibfnamefont {W.~R.}\ \bibnamefont {Milner}},
  \bibinfo {author} {\bibfnamefont {J.}~\bibnamefont {De~Hond}}, \bibinfo
  {author} {\bibfnamefont {J.~F.}\ \bibnamefont {Fricke}},\ and\ \bibinfo
  {author} {\bibfnamefont {W.}~\bibnamefont {Ketterle}},\ }\href
  {https://doi.org/10.1103/PhysRevLett.134.183401} {\bibfield  {journal}
  {\bibinfo  {journal} {Physical Review Letters}\ }\textbf {\bibinfo {volume}
  {134}},\ \bibinfo {pages} {183401} (\bibinfo {year} {2025})}\BibitemShut
  {NoStop}%
\bibitem [{\citenamefont {Dixmerias}\ \emph {et~al.}(2025)\citenamefont
  {Dixmerias}, \citenamefont {Vecchio}, \citenamefont {Daix}, \citenamefont
  {Verstraten}, \citenamefont {de~Jongh}, \citenamefont {Peaudecerf},
  \citenamefont {Doussal}, \citenamefont {Schehr},\ and\ \citenamefont
  {Yefsah}}]{dixmerias2025a}%
  \BibitemOpen
  \bibfield  {author} {\bibinfo {author} {\bibfnamefont {M.}~\bibnamefont
  {Dixmerias}}, \bibinfo {author} {\bibfnamefont {G.~D. V.~D.}\ \bibnamefont
  {Vecchio}}, \bibinfo {author} {\bibfnamefont {C.}~\bibnamefont {Daix}},
  \bibinfo {author} {\bibfnamefont {J.}~\bibnamefont {Verstraten}}, \bibinfo
  {author} {\bibfnamefont {T.}~\bibnamefont {de~Jongh}}, \bibinfo {author}
  {\bibfnamefont {B.}~\bibnamefont {Peaudecerf}}, \bibinfo {author}
  {\bibfnamefont {P.~L.}\ \bibnamefont {Doussal}}, \bibinfo {author}
  {\bibfnamefont {G.}~\bibnamefont {Schehr}},\ and\ \bibinfo {author}
  {\bibfnamefont {T.}~\bibnamefont {Yefsah}},\ }\href
  {https://doi.org/10.48550/arXiv.2510.25735} {\bibinfo {title} {Universal
  {{Random Matrix Behavior}} of a {{Fermionic Quantum Gas}}}} (\bibinfo {year}
  {2025}),\ \Eprint {https://arxiv.org/abs/2510.25735} {arXiv:2510.25735
  [cond-mat]} \BibitemShut {NoStop}%
\bibitem [{\citenamefont {{Obeso-Jureidini}}\ and\ \citenamefont
  {{Romero-Roch{\'i}n}}(2022)}]{obeso-jureidini2022}%
  \BibitemOpen
  \bibfield  {author} {\bibinfo {author} {\bibfnamefont {J.~C.}\ \bibnamefont
  {{Obeso-Jureidini}}}\ and\ \bibinfo {author} {\bibfnamefont {V.}~\bibnamefont
  {{Romero-Roch{\'i}n}}},\ }\href {https://doi.org/10.1103/PhysRevA.105.043307}
  {\bibfield  {journal} {\bibinfo  {journal} {Physical Review A}\ }\textbf
  {\bibinfo {volume} {105}},\ \bibinfo {pages} {043307} (\bibinfo {year}
  {2022})}\BibitemShut {NoStop}%
\bibitem [{\citenamefont {Schweigler}\ \emph {et~al.}(2017)\citenamefont
  {Schweigler}, \citenamefont {Kasper}, \citenamefont {Erne}, \citenamefont
  {Mazets}, \citenamefont {Rauer}, \citenamefont {Cataldini}, \citenamefont
  {Langen}, \citenamefont {Gasenzer}, \citenamefont {Berges},\ and\
  \citenamefont {Schmiedmayer}}]{schweigler2017}%
  \BibitemOpen
  \bibfield  {author} {\bibinfo {author} {\bibfnamefont {T.}~\bibnamefont
  {Schweigler}}, \bibinfo {author} {\bibfnamefont {V.}~\bibnamefont {Kasper}},
  \bibinfo {author} {\bibfnamefont {S.}~\bibnamefont {Erne}}, \bibinfo {author}
  {\bibfnamefont {I.}~\bibnamefont {Mazets}}, \bibinfo {author} {\bibfnamefont
  {B.}~\bibnamefont {Rauer}}, \bibinfo {author} {\bibfnamefont
  {F.}~\bibnamefont {Cataldini}}, \bibinfo {author} {\bibfnamefont
  {T.}~\bibnamefont {Langen}}, \bibinfo {author} {\bibfnamefont
  {T.}~\bibnamefont {Gasenzer}}, \bibinfo {author} {\bibfnamefont
  {J.}~\bibnamefont {Berges}},\ and\ \bibinfo {author} {\bibfnamefont
  {J.}~\bibnamefont {Schmiedmayer}},\ }\href
  {https://doi.org/10.1038/nature22310} {\bibfield  {journal} {\bibinfo
  {journal} {Nature}\ }\textbf {\bibinfo {volume} {545}},\ \bibinfo {pages}
  {323} (\bibinfo {year} {2017})}\BibitemShut {NoStop}%
\bibitem [{\citenamefont {Rispoli}\ \emph {et~al.}(2019)\citenamefont
  {Rispoli}, \citenamefont {Lukin}, \citenamefont {Schittko}, \citenamefont
  {Kim}, \citenamefont {Tai}, \citenamefont {Léonard},\ and\ \citenamefont
  {Greiner}}]{rispoli2019}%
  \BibitemOpen
  \bibfield  {author} {\bibinfo {author} {\bibfnamefont {M.}~\bibnamefont
  {Rispoli}}, \bibinfo {author} {\bibfnamefont {A.}~\bibnamefont {Lukin}},
  \bibinfo {author} {\bibfnamefont {R.}~\bibnamefont {Schittko}}, \bibinfo
  {author} {\bibfnamefont {S.}~\bibnamefont {Kim}}, \bibinfo {author}
  {\bibfnamefont {M.~E.}\ \bibnamefont {Tai}}, \bibinfo {author} {\bibfnamefont
  {J.}~\bibnamefont {Léonard}},\ and\ \bibinfo {author} {\bibfnamefont
  {M.}~\bibnamefont {Greiner}},\ }\href
  {https://doi.org/10.1038/s41586-019-1527-2} {\bibfield  {journal} {\bibinfo
  {journal} {Nature}\ }\textbf {\bibinfo {volume} {573}},\ \bibinfo {pages}
  {385} (\bibinfo {year} {2019})}\BibitemShut {NoStop}%
\bibitem [{\citenamefont {Christakis}\ \emph {et~al.}(2023)\citenamefont
  {Christakis}, \citenamefont {Rosenberg}, \citenamefont {Raj}, \citenamefont
  {Chi}, \citenamefont {Morningstar}, \citenamefont {Huse}, \citenamefont
  {Yan},\ and\ \citenamefont {Bakr}}]{christakis2023}%
  \BibitemOpen
  \bibfield  {author} {\bibinfo {author} {\bibfnamefont {L.}~\bibnamefont
  {Christakis}}, \bibinfo {author} {\bibfnamefont {J.~S.}\ \bibnamefont
  {Rosenberg}}, \bibinfo {author} {\bibfnamefont {R.}~\bibnamefont {Raj}},
  \bibinfo {author} {\bibfnamefont {S.}~\bibnamefont {Chi}}, \bibinfo {author}
  {\bibfnamefont {A.}~\bibnamefont {Morningstar}}, \bibinfo {author}
  {\bibfnamefont {D.~A.}\ \bibnamefont {Huse}}, \bibinfo {author}
  {\bibfnamefont {Z.~Z.}\ \bibnamefont {Yan}},\ and\ \bibinfo {author}
  {\bibfnamefont {W.~S.}\ \bibnamefont {Bakr}},\ }\href
  {https://doi.org/10.1038/s41586-022-05558-4} {\bibfield  {journal} {\bibinfo
  {journal} {Nature}\ }\textbf {\bibinfo {volume} {614}},\ \bibinfo {pages}
  {64} (\bibinfo {year} {2023})}\BibitemShut {NoStop}%
\bibitem [{\citenamefont {Bureik}\ \emph {et~al.}(2025)\citenamefont {Bureik},
  \citenamefont {Hercé}, \citenamefont {Allemand}, \citenamefont {Tenart},
  \citenamefont {Roscilde},\ and\ \citenamefont {Clément}}]{bureik2025}%
  \BibitemOpen
  \bibfield  {author} {\bibinfo {author} {\bibfnamefont {J.-P.}\ \bibnamefont
  {Bureik}}, \bibinfo {author} {\bibfnamefont {G.}~\bibnamefont {Hercé}},
  \bibinfo {author} {\bibfnamefont {M.}~\bibnamefont {Allemand}}, \bibinfo
  {author} {\bibfnamefont {A.}~\bibnamefont {Tenart}}, \bibinfo {author}
  {\bibfnamefont {T.}~\bibnamefont {Roscilde}},\ and\ \bibinfo {author}
  {\bibfnamefont {D.}~\bibnamefont {Clément}},\ }\href
  {https://doi.org/10.1038/s41567-024-02700-z} {\bibfield  {journal} {\bibinfo
  {journal} {Nature Physics}\ }\textbf {\bibinfo {volume} {21}},\ \bibinfo
  {pages} {57} (\bibinfo {year} {2025})}\BibitemShut {NoStop}%
\bibitem [{\citenamefont {Allemand}\ \emph {et~al.}(2025)\citenamefont
  {Allemand}, \citenamefont {Dupuy}, \citenamefont {Paquiez}, \citenamefont
  {Dupuis}, \citenamefont {Rançon}, \citenamefont {Roscilde}, \citenamefont
  {Chalopin},\ and\ \citenamefont {Clément}}]{allemand2025}%
  \BibitemOpen
  \bibfield  {author} {\bibinfo {author} {\bibfnamefont {M.}~\bibnamefont
  {Allemand}}, \bibinfo {author} {\bibfnamefont {G.}~\bibnamefont {Dupuy}},
  \bibinfo {author} {\bibfnamefont {P.}~\bibnamefont {Paquiez}}, \bibinfo
  {author} {\bibfnamefont {N.}~\bibnamefont {Dupuis}}, \bibinfo {author}
  {\bibfnamefont {A.}~\bibnamefont {Rançon}}, \bibinfo {author} {\bibfnamefont
  {T.}~\bibnamefont {Roscilde}}, \bibinfo {author} {\bibfnamefont
  {T.}~\bibnamefont {Chalopin}},\ and\ \bibinfo {author} {\bibfnamefont
  {D.}~\bibnamefont {Clément}},\ }\href
  {https://doi.org/10.48550/arXiv.2508.21623} {\bibinfo {title} {Observation of
  universal non-{Gaussian} statistics of the order parameter across a
  continuous phase transition}} (\bibinfo {year} {2025}),\ \bibinfo {note}
  {arXiv:2508.21623 [cond-mat]}\BibitemShut {NoStop}%
\bibitem [{\citenamefont {Jin}\ \emph {et~al.}(2024)\citenamefont {Jin},
  \citenamefont {Dai}, \citenamefont {Verstraten}, \citenamefont {Dixmerias},
  \citenamefont {Alhyder}, \citenamefont {Salomon}, \citenamefont {Peaudecerf},
  \citenamefont {{de Jongh}},\ and\ \citenamefont {Yefsah}}]{jin2024}%
  \BibitemOpen
  \bibfield  {author} {\bibinfo {author} {\bibfnamefont {S.}~\bibnamefont
  {Jin}}, \bibinfo {author} {\bibfnamefont {K.}~\bibnamefont {Dai}}, \bibinfo
  {author} {\bibfnamefont {J.}~\bibnamefont {Verstraten}}, \bibinfo {author}
  {\bibfnamefont {M.}~\bibnamefont {Dixmerias}}, \bibinfo {author}
  {\bibfnamefont {R.}~\bibnamefont {Alhyder}}, \bibinfo {author} {\bibfnamefont
  {C.}~\bibnamefont {Salomon}}, \bibinfo {author} {\bibfnamefont
  {B.}~\bibnamefont {Peaudecerf}}, \bibinfo {author} {\bibfnamefont
  {T.}~\bibnamefont {{de Jongh}}},\ and\ \bibinfo {author} {\bibfnamefont
  {T.}~\bibnamefont {Yefsah}},\ }\href
  {https://doi.org/10.1103/PhysRevResearch.6.013158} {\bibfield  {journal}
  {\bibinfo  {journal} {Physical Review Research}\ }\textbf {\bibinfo {volume}
  {6}},\ \bibinfo {pages} {013158} (\bibinfo {year} {2024})}\BibitemShut
  {NoStop}%
\end{thebibliography}
\end{document}